\documentclass[sigconf]{aamas}

\usepackage{balance} 

\setcopyright{ifaamas}
\acmConference[AAMAS '25]{Under review of the 24th International Conference
on Autonomous Agents and Multiagent Systems (AAMAS 2025)}{May 19 -- 23, 2025}
{Detroit, Michigan, USA}{A.~El~Fallah~Seghrouchni, Y.~Vorobeychik, S.~Das, A.~Nowe (eds.)}
\copyrightyear{2025}
\acmYear{2025}
\acmDOI{}
\acmPrice{}
\acmISBN{}

\title{IBGP: Imperfect Byzantine Generals Problem for Zero-Shot Robustness in Communicative Multi-Agent Systems}

\author{Yihuan Mao\textsuperscript{*}}
\affiliation{
  \institution{Tsinghua University}
  \city{Beijing}
  \country{China}
}
\email{maoyh20@mails.tsinghua.edu.cn}

\author{Yipeng Kang\textsuperscript{*}}
\affiliation{
  \institution{State Key Laboratory of General Artificial Intelligence, BIGAI}
  \city{Beijing}
  \country{China}
}
\email{kangyipeng@bigai.ai}

\author{Peilun Li}
\affiliation{
  \institution{Shanghai Tree-Graph Blockchain Research Institute}
  \city{Shanghai}
  \country{China}
}
\email{peilun.li@confluxnetwork.org}

\author{Ning Zhang}
\affiliation{
  \institution{Washington University in St. Louis}
  \city{St. Louis, 63130}
  \country{United States}}
\email{zhang.ning@wustl.edu}

\author{Wei Xu}
\affiliation{
  \institution{Tsinghua University}
  \city{Beijing}
  \country{China}}
\email{weixu@tsinghua.edu.cn}

\author{Chongjie Zhang}
\affiliation{
  \institution{Washington University in St. Louis}
  \city{St. Louis, 63130}
  \country{United States}}
\email{chongjie@wustl.edu}

\usepackage[utf8]{inputenc} 
\usepackage[T1]{fontenc}    
\usepackage{hyperref}       
\usepackage{url}            
\usepackage{booktabs}       
\usepackage{amsfonts}       
\usepackage{nicefrac}       
\usepackage{microtype}      
\usepackage{xcolor}         

\usepackage{soul}
\usepackage{url}
\usepackage{graphicx}
\usepackage{amsmath}
\usepackage{amsthm}
\usepackage{booktabs}
\usepackage{bbm}
\usepackage{xcolor}
\usepackage{multirow}
\usepackage[ruled]{algorithm2e}
\usepackage{subfig}
\usepackage{float}
\usepackage{bm}
\usepackage{amsthm}
\usepackage{amsmath}
\usepackage{makecell}
\usepackage{array}
\usepackage{colortbl}
\usepackage{dsfont}

\newtheorem{theorem}{Theorem}
\newtheorem{lemma}{Lemma}
\newtheorem{definition}{Definition}

\begin{abstract}

As large language model (LLM) agents increasingly integrate into our infrastructure, their robust coordination and message synchronization become vital. The Byzantine Generals Problem (BGP) is a critical model for constructing resilient multi-agent systems (MAS) under adversarial attacks. It describes a scenario where malicious agents with unknown identities exist in the system-situations that, in our context, could result from LLM agents' hallucinations or external attacks. In BGP, the objective of the entire system is to reach a consensus on the action to be taken. Traditional BGP requires global consensus among all agents; however, in practical scenarios, global consensus is not always necessary and can even be inefficient. Therefore, there is a pressing need to explore a refined version of BGP that aligns with the local coordination patterns observed in MAS. We refer to this refined version as Imperfect BGP (IBGP) in our research, aiming to address this discrepancy.

To tackle this issue, we propose a framework that leverages consensus protocols within general MAS settings, providing provable resilience against communication attacks and adaptability to changing environments, as validated by empirical results. Additionally, we present a case study in a sensor network environment to illustrate the practical application of our protocol.
\end{abstract}


\begin{document}

\pagestyle{fancy}
\fancyhead{}

\maketitle

\renewcommand\thefootnote{\fnsymbol{footnote}} 
\footnotetext[1]{Equal contribution.}

\section{Introduction}
With the advancement of AI technology, the world is entering a new era in which AI agents will constitute a significant portion of our infrastructure. In the foreseeable future, spontaneously assembled groups of heterogeneous AI agents—of various types and functions, developed by different manufacturers—will frequently interact to solve temporary daily tasks. The coexistence and coordination of diverse agents will be crucial, much like traditional human coordination.

A key category of coordination involves the synchronization of messages among agents. This is exemplified in applications such as sensor networks \cite{olfati2004consensus} and UAV control \cite{dorri2018multi}. When the agents are heterogeneous, such as autonomous vehicles \cite{kober2021smarts} with communication modules driven by general-purpose large language models, they are not specifically designed for message synchronization. Therefore, we can reasonably assume that agents lack a reliable predefined broadcast module to ensure consistent messaging during synchronization. Discrepancies may arise between their messages and final actions for two reasons: (1) the agents' outputs may be affected by hallucinations, and (2) agents may be compromised and act maliciously. Overall, the messages and actions of general-purpose agents are often based on natural language rather than predefined protocols, making abnormal behavior likely during synchronization tasks. Malicious agents can lead to disastrous consequences; for instance, tampering with vehicle-to-vehicle (V2V) messages can result in miscoordination, potentially causing significant property damage and loss of life. This security concern extends beyond autonomous vehicles to various applications requiring reliable coordination. For instance, various applications necessitate the coordination of agents with different functionalities, such as in video generation \citep{Wu_Huang_Yu_2024}, software development \citep{hong2023metagpt}, and problem-solving \citep{wu2023autogen}.

Traditionally, this issue relates to the consensus problem in distributed systems, such as the Byzantine Generals Problem (BGP) \cite{lamport1982byzantine}, where the primary objective is to synchronize content across all benign nodes and achieve system-wide consistency, even in the presence of node failures. In this context, a perfect protocol is essential. However, in the coordination of multi-agent systems, consensus serves as a means to achieve team goals, allowing for a relaxation of stringent requirements; often, it suffices to synchronize only a minimal number of benign agents.

For example, in a Predator-Prey environment \cite{son2019qtran}, where predators need to cooperate to hunt, miscoordination could be fatal, as a single predator might be killed by the prey. However, a global consensus is unnecessary, as not all predators need to participate in the hunt. Another example is the threshold public goods game \cite{daiki2020public} in experimental economics, which mirrors global cooperation on climate or energy issues: agents contribute funds to a shared pot, and positive utility is only achieved when contributions exceed a threshold, necessitating close collaboration.

\begin{figure*}[ht]
\centering
\includegraphics[width=0.95\textwidth]{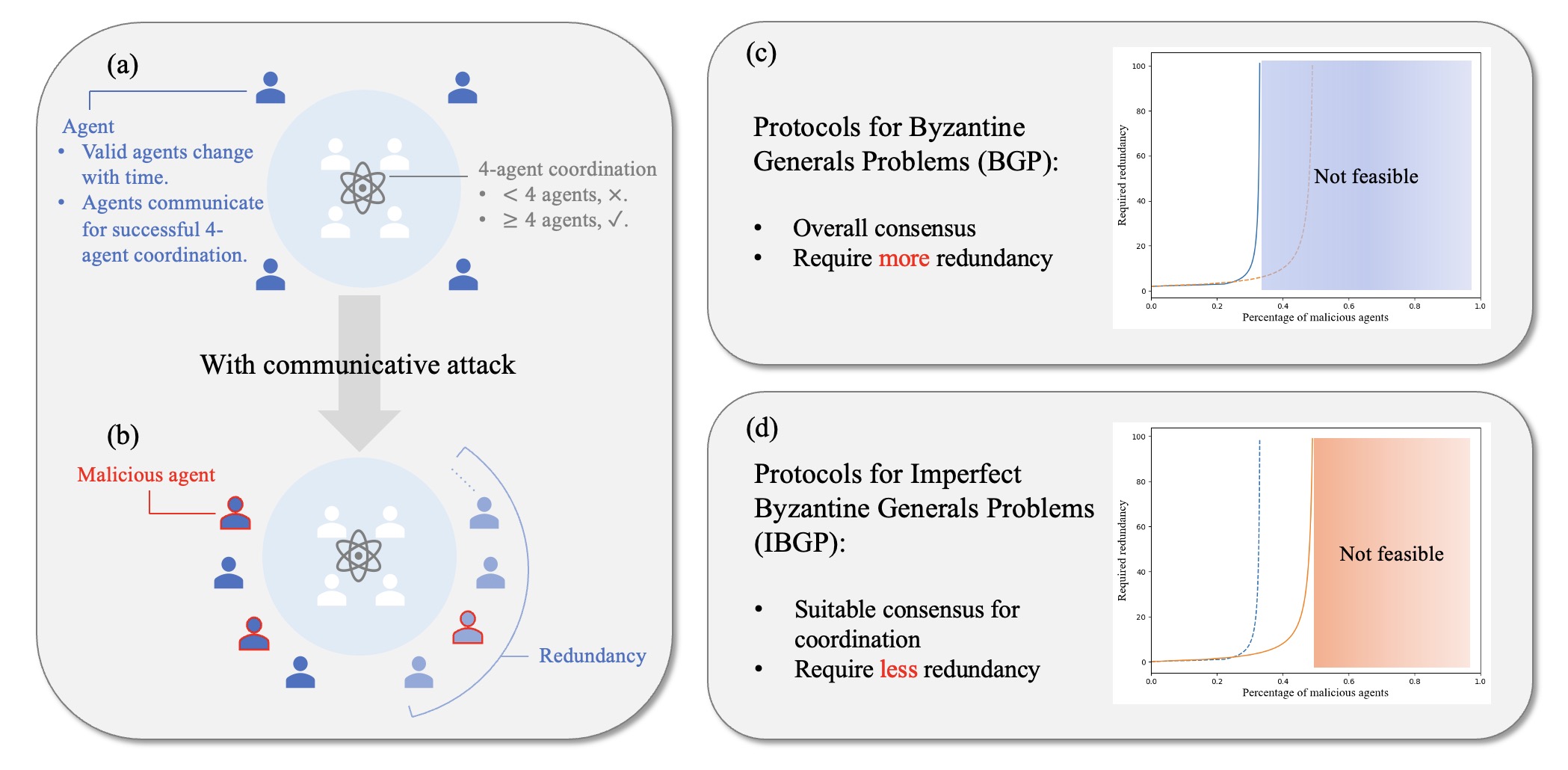}
\caption{An illustration of the motivation for proposing IBGP. (a) Agents coordinate on a task that requires four participants. (b) The presence of malicious agents capable of sending misleading messages may lead to the failure of coordination among the original four agents. As a result, some redundancy is necessary to ensure safe coordination through consensus protocols. (c) Traditional methods for achieving overall consensus involve developing BGP protocols, but these are inefficient and require excessive redundancy in coordination contexts because they're specifically designed for overall consensus scenarios. The graph shows that BGP becomes infeasible when malicious agents exceed 33\%.(d) We define the coordination problem under communicative attack as IBGP and propose tailored protocols for it. Our protocol is suitable for coordination settings and requires less redundancy, accommodating a higher percentage of malicious agents (up to 50\%). Details about (c) and (d) is provided in Appendix \ref{intro_fig_detail}.} \label{fig:intro} \end{figure*}

To tackle the coordination challenges in multi-agent systems, we first formalize the issue, referring to it as the Imperfect Byzantine Generals Problem (IBGP). While it shares similarities with the classic BGP, IBGP fundamentally differs in its coordination requirements, emphasizing partial consensus. We then introduce a protocol specifically designed for the IBGP. As shown in Figure \ref{fig:intro}, our IBGP protocol demands less redundancy and accommodates a higher percentage of malicious agents compared to existing BGP protocols.

In the realm of AGI, the implementation of our protocol does not require additional development steps; instead, it can be integrated into agents through simple prompt instructions, leveraging the agents' inherent capabilities. If an agent fails to understand or execute this protocol, or if it intentionally disrupts the process following an attack, it will be classified as a malicious node. As long as the number of such nodes remains low, our protocol can continue to function effectively. Experiments demonstrate the effectiveness of our consensus protocols in both IBGP and more practical tasks.

Our technical contributions are two-fold: (i) we define IBGP, highlighting the challenges of partial coordination in Multi-Agent Systems, and (ii) we devise an innovative consensus protocol that ensures secure coordination, validated through theoretical analysis and practical experimentation.

\begin{table*}
    \centering
    \begin{tabular}{c|c}
        \toprule
        \multicolumn{2}{c}{$M^0_i\in\{0,1\},a_i\in\{0,1\}$}\\
        \midrule
        BGP& IBGP\\
        \midrule
        $n$ benign agents &$n$ benign agents, coordination threshold $k$\\
        \midrule
        $
        R = 
        \left\{
        \begin{array}{ll}
            \mathds{1}(\#(a_i=1)=n) & \hspace{-2.5em}\text{if } \#(M^0_i=1)=n, \\
            \mathds{1}(\#(a_i=1)=0) & \hspace{-2.5em}\text{if } \#(M^0_i=1)=0, \\
            \mathds{1}(\#(a_i=1)=n) + \mathds{1}(\#(a_i=1)=0) & \hspace{0em}\text{otherwise}.
        \end{array}
        \right.
        $
        &
        $
        R =
        \left\{
        \begin{array}{ll}
            \mathds{1}(\#(M^0_i=1,a_i=1)\ge k) & \hspace{-3em}\text{if }\ \#(M^0_i=1)=n,\\
            \mathds{1}(\#(M^0_i=1,a_i=1)=0) & \hspace{-3em}\text{if }\ \#(M^0_i=1)<k,\\
            \mathds{1}(\#(M^0_i=1,a_i=1)\ge k) + \mathbbm{1}(\#(M^0_i=1,a_i=1)=0) & \hspace{0em} \text{otherwise.}
        \end{array}
        \right.
        $\\
        \bottomrule
    \end{tabular}
    \caption{The difference of definition between BGP and IBGP.}
    \label{tab:bgp}
\end{table*}

\section{Related Work}
The Byzantine Generals Problem \cite{lamport1982byzantine} describes a scenario where a group of Byzantine generals must agree on a common plan of action, even though some of the generals may be traitors. This models a large type of distributed consistency problem in computer science. Byzantine consensus protocols in a distributed network can be used to reach an agreement on a single value, even in the presence of faulty or malicious nodes. Important works of Byzantine consensus protocols include Practical Byzantine Fault Tolerance (PBFT) \cite{castro1999practical} and Randomized Byzantine Generals \cite{rabin1983randomized}. PBFT is a solution to the Byzantine Generals Problem that is designed for practical use in distributed systems. Rabin's work \cite{rabin1983randomized} provided a solution to the Byzantine Generals Problem that did not require a centralized authority or a trusted third party. Instead, it introduced the idea of a randomized protocol where each node chooses a random value that is used to break ties in the event of conflicting messages. The randomization ensures that Byzantine nodes cannot predict the outcome of the protocol, making it more difficult for them to interfere with the consensus process.

In the context of MAS research, one important research branch of MAS involves learning a communications system to achieve a common goal. Some use explicit communication systems with discrete or continuous signals, mainly to convey informative local observations to each other, to deal with partial observation \cite{tonghan2020learning,kang2020incorporating}. Some use communication for global value optimization \cite{bohmer2020deep,kang2022non}. 
Different from these categories, in MAS research, consensus refers to achieving a global agreement over a particular feature of interest \cite{dorri2018multi,yu2009second,fu2014adaptive,olfati2004consensus,olfati2006flocking,fan2014bipartite,liu2014containment}. It has been widely studied as it affects communication and collaboration between agents. However, very few of them paid attention to adversarial attacks. Recently some researchers have investigated algorithms to mitigate the impact of malicious agents \cite{sun2023certifiably,xue2022mis}. However, their methodologies lack zero-shot adaption ability to the attackers and the environment.

\section{Imperfect Byzantine Generals Problem (IBGP)}
\label{sec:problem}

\subsection{Preliminaries: BGP}
\label{bgp}
In BGP, there are $n$ benign agents and $t$ attacker agents communicating with each other. The attackers can send false messages to disturb coordination. Each agent begins with an initial message $M^0\in\{0,1\}$ as its initial proposal and finally makes a decision action $a\in\{0,1\}$. Generally, a value of $1$ indicates cooperation, while $0$ indicates giving up. The formal definition of BGP is provided in Definition \ref{def:bgp}. For clarity, we also present an equivalent definition within the context of Reinforcement Learning, as depicted by the reward function in the left portion of Table \ref{tab:bgp}, which outlines the optimal action based on certain observations.

\begin{definition}
\label{def:bgp}
A system solves BGP if it meets the following requirements under any attack:
\begin{itemize}
    \item \textit{Agreement:} $a_1 = a_2 = \cdots = a_n \in \{0, 1\}$. (All $n$ benign agents must agree on the same action, either $0$ or $1$.)
    \item \textit{Consistency:} If $M^0_1 = M^0_2 = \cdots = M^0_n = x$, then $a_1 = a_2 = \cdots = a_n = x$. (When all $n$ agents have identical initial observations, their actions must also be identical.)
\end{itemize}
\end{definition}

All the agents, including both the agents and the attackers, can communicate through a complete network for a few rounds. Agents decide what messages to send at the end of each round after reading messages from other agents. In BGP, the attackers aim to break the consensus of agents, but the agents are unaware of the identity of the communicating agents (benign or malicious).

Mis-coordination describes the situation where \textit{Agreement} is violated, and researchers have designed consensus protocols proven to prevent mis-coordination under any communicative attacks.

\begin{figure*}
\label{pipeline}
\centering
\includegraphics[width=0.95\textwidth]{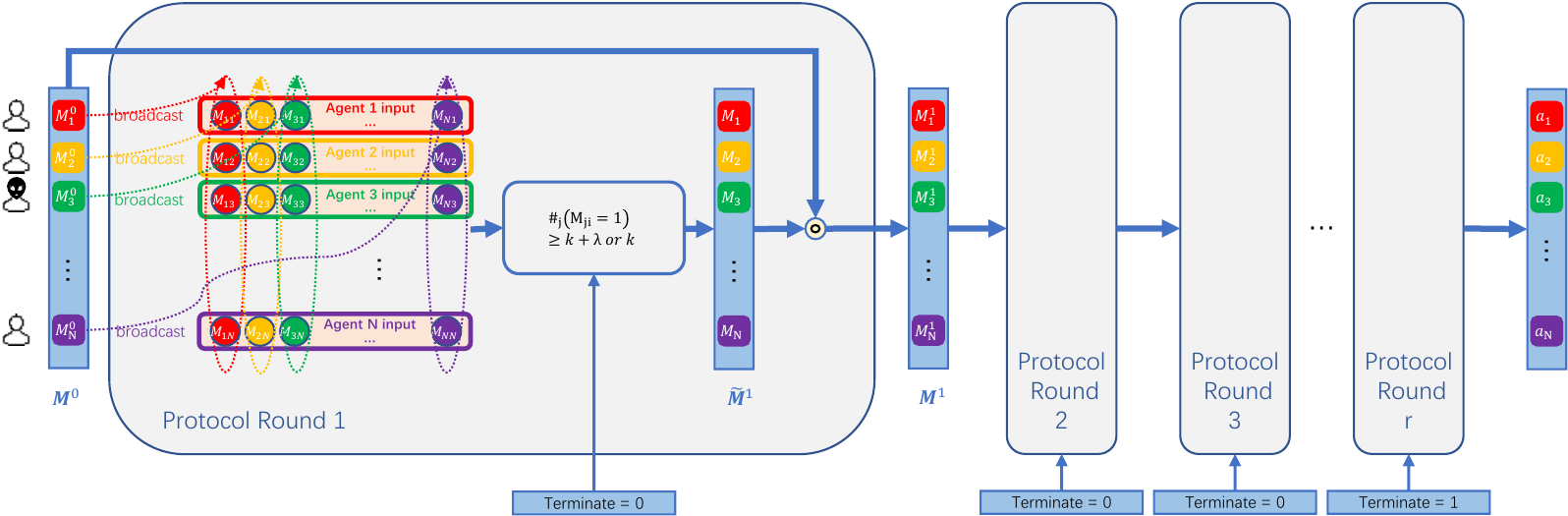}
\caption{A general framework for consensus protocol.}
\label{framework}
\end{figure*}

\subsection{IBGP}
\label{ibgp_subsection}
BGP serves as a fundamental concept of extensive research within the area of Decentralized Systems, embodying the crucial attributes of agreement and consistency within a decentralized framework. Nevertheless, these properties may not always apply in many Multi-Agent Systems (MAS), where partial coordination is a common pattern rather than universal coordination. For example, in a predator-prey environment \cite{stone2000multiagent}, only a subset of predators may be required to collaborate in hunting a particular prey, rather than involving all predators in the pursuit.

To capture this coordination pattern within MAS, we introduce IBGP in Definition \ref{def:ibgp}, where successful agreement necessitates the cooperation of only $k$. Besides, only the agents with the initial observation $M^0=1$ are permitted to take the cooperative action $a=1$. The two properties of \textit{Agreement} and \textit{Consistency} are redefined to align with the partial coordination prevalent in Multi-Agent Systems. Table \ref{tab:bgp} illustrates the distinctions between IBGP and the original BGP.

\begin{definition}
\label{def:ibgp}
A system solves IBGP if if it meets the following requirements under any attack:
\begin{itemize}
    \item \textit{Agreement:} $\#(M^0_i = 1, a_i = 1) \in \{0\} \cup [k, n]$. (At least $k$ agents that observe $M^0=1$ are required to cooperate; otherwise no agent should act.)
    \item \textit{Consistency:} If $\#(M^0_i = 1) = n$, then $\#(M^0_i = 1, a_i = 1) \geq k$. (If the number of available agents is super-sufficient, cooperation must happen.)
\end{itemize}
\end{definition}

Similarly, mis-coordination is defined as the situation $0<\#(M^0_i=1,a_i=1)<k$, where \textit{Agreement} is violated. It means that some agents try to coordinate but fail. Just like BGP, the goal of IBGP is to avoid mis-coordination altogether, and although we use Reinforcement Learning to formulate BGP and IBGP in the following section, our focus is on robustness rather than expected return.


\begin{figure}
    \centering
        \subfloat[An example of mis-coordination when $\lambda=4$.]{\label{img:ibgp_l4}
        \includegraphics[width=0.9\linewidth]{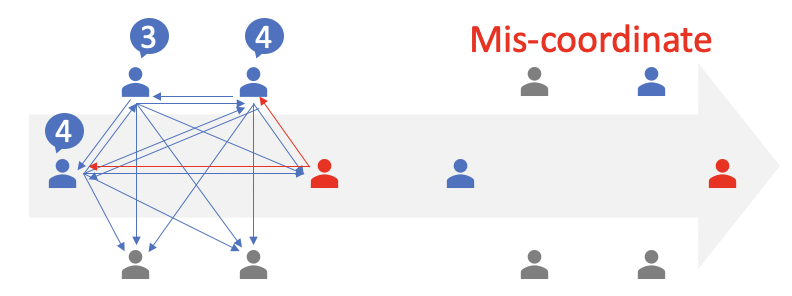}
        }\\
        \subfloat[An example of mis-coordination when $\lambda=5$.]{\label{img:ibgp_l5}
        \includegraphics[width=0.9\linewidth]{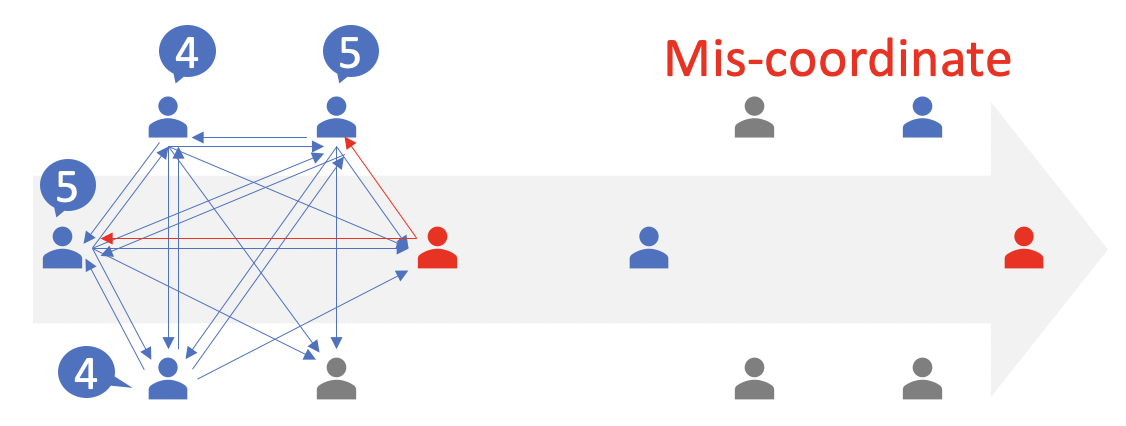}
        }
    \caption{Examples of mis-coordination under different $\lambda$s, showing the limitation of single-round decision process.}
    \label{ibgp_challenge}
\end{figure}

\paragraph{\bf Didactic example} To understand the challenge of IBGP, we illustrate that a single-round decision process is inadequate for addressing such issues through a didactic example featuring a specific attacking pattern.

First, it's obvious that a reasonable single-round decision process for agent $i$ is to take $a_i=1$ if the number of received signals exceeds a given threshold, i.e., $\#(M_{\cdot\rightarrow i}=1)\ge\lambda$, where $\lambda$ is the threshold of the decision process. For simplicity, we assume the strategy shares with all agents.

In Figure \ref{ibgp_challenge}, the IBGP with $n=5,t=1,k=3$ includes $5$ benign agents and $1$ attacker. Figure \ref{img:ibgp_l4} shows an example of mis-coordination when the threshold of the single-round process is set to $\lambda=4$. The number of initial active honest agents $\#(M^0_i=1)=3$. After sending messages, agents sum up the received messages, and some get $3$ messages while others get $4$ messages. This results in that the agent who receives $3$ messages give up and causes mis-coordination. The example in Figure \ref{img:ibgp_l5} shows that mis-coordination could still happen even if we raise the threshold. When the threshold is set to $\lambda=5$, the failed example happens in a different initialization, where there are $4$ active honest agents. After sending messages, the sum of received messages are $4$ and $5$ respectively. Since the agents receiving $4$ messages give up and the agents receiving $5$ messages still try to coordinate, the mis-coordination happens again.

\begin{figure*}
    \centering
        \subfloat[When the number of initial active honest agents $\#(M^0_i=1)=3$, the iteration doesn't cause final mis-coordination.]{\label{img:con_0}
        \includegraphics[width=0.9\linewidth]{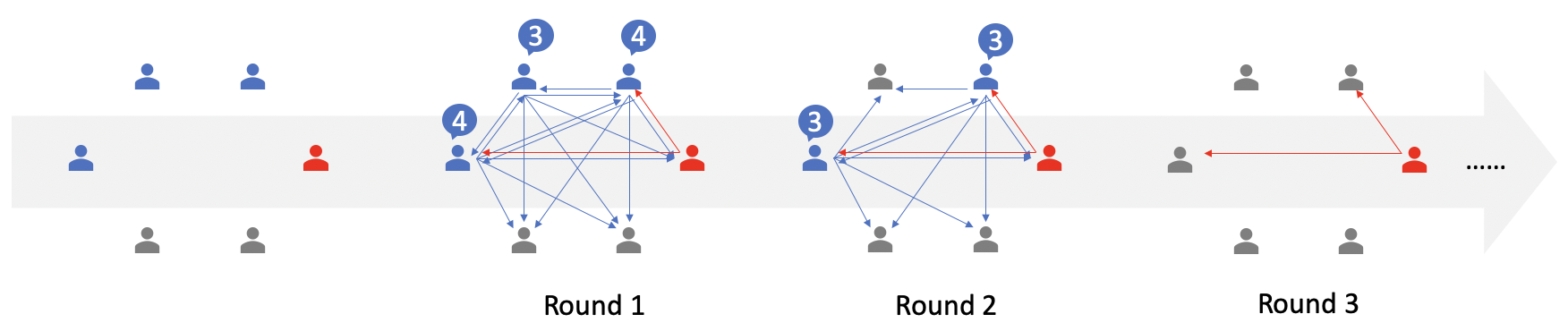}
        }
        \quad
        \subfloat[When the number of initial active honest agents $\#(M^0_i=1)=4$, the iteration doesn't cause final mis-coordination.]{\label{img:con_1}
        \includegraphics[width=0.9\linewidth]{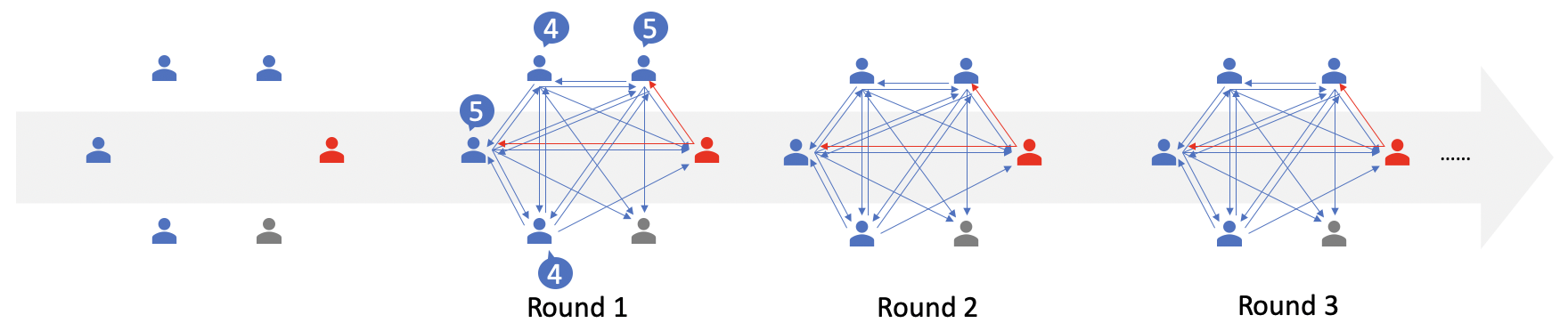}
        }
    \caption{Examples of the iterations of the protocol.}
    \label{img:did_protocol}
\end{figure*}

\section{Method}
\subsection{Consensus Protocol for IBGP}
\label{prot}

To solve IBGP, we propose a consensus protocol as illustrated in Figure \ref{framework}. This protocol employs a multi-round broadcast pattern and incorporates the concept of an independent global randomizer (implemented similarly with \cite{rabin1983randomized}). In each round, a randomized bit variable determines whether it is the last round (with a value of 1 indicating the final round and 0 indicating that the process should continue). Agents broadcast their proposals and deal with the received proposals under the effect of the randomized bit. The results serve as the proposals for the next round or as the final decisions in the last round. The default proposals and decisions are both initialized to 0 if not specified. The framework amounts to the $(k,\lambda)$-protocol listed below:

\begin{enumerate}
    \item The global randomizer initializes the number of rounds from a distribution $r_{tot}\sim\mathcal{R}$.
            ($\mathcal{R}$ is the sample distribution of the total number of rounds $r_{tot}$ in the IBGP Protocol. $r_{tot}$ isn't revealed until the last round arrives.)
    \item Initial round: Each agent $i$ broadcasts its initial proposal $M_{i}^{0}$ to each agents $j$.
    \item Round $r\in\{1\cdots r_{tot}\}$: Each agent $i\in\{i|M^{r-1}_i=1\}$ broadcasts $M^{r}_{i}=\mathbbm{1}(\#_{j\in[N]}(M^{r-1}_{j\to i}=1)\ge k+\lambda)$.
    \item Decision making round: Each agent $i\in\{i|M^{r_{tot}}_i=1\}$ select action $a_i=\mathbbm{1}(\#_{j\in[N]}(M^{r_{tot}}_{j\to i}=1)\ge k)$.
\end{enumerate}

When assigning $\lambda=t$, we have the following theorem with its proof in Appendix \ref{ibgp_proof}:
\begin{theorem}
\label{thm:protocol}
    $(k,t)$-protocol is robust with a high level of confidence $1-\max_{r}\{p(r_{tot}=r)\}$ under any attack on IBGP(t, k).
\end{theorem}

Note that the iterative rounds of IBGP communication won't bottleneck the efficiency of our system, because of the simplicity of the communication computation. There's just basic pairwise bit communication among agents and a straightforward linear decision-making process for each agent. This simplicity stands in contrast to the more complex processes found elsewhere in the MARL system. Additionally, in real-world multi-agent communication contexts, using multiple rounds of communication for sending one piece of content is typical and the cost is affordable.

\paragraph{\bf Didactic example} We provide a didactic example to illustrate how the protocol operates. Proving the robustness of a method requires considering all potential scenarios, which can be space-intensive, or relying on the mathematical proof of Theorem \ref{thm:protocol}. Therefore, we focus on how the protocol prevents the mis-coordination shown in the didactic example of IBGP in Section \ref{ibgp_subsection}. A comprehensive exploration of all possible scenarios is discussed in Appendix \ref{diagram}.

The following game is an IBGP with $n=5,t=1,k=3$, including $5$ benign agents and $1$ attacker, and we show how the $(k,t)$-protocol works under different initializations. In Figure \ref{img:con_0}, the intial number of active honest agents $\#(M^0_i=1)=3$. The threshold of a single round is dynamically decided by the global randomizer (from $3$ and $4$), and therefore the agent receiving $3$ messages is probably both to continue or give up. If it continues, the active agents remains the same. If it give up (the third subfigure), the number of active agents decreases to $2$, which is a temporary mis-coordination. Luckily, such temporary mis-coordination only lasts for $1$ round, and in the next round, all agents will give up due to the fewer received messages. In conclusion, under the iterative protocol, the temporary mis-coordination can only exists for a single round, and since the real number of rounds is unknown to the attacker, the successful attack can only happen with a low probability. In Figure \ref{img:con_1}, since the threshold is $3$ or $4$, they will always coordinate.

\subsection{Integrating IBGP with Multi-Agent Reinforcement Learning}
\label{pipeline_subsection}
To manifest the effectiveness of the IBGP consensus protocol, we invoke it as a coordination module in the decentralized multi-agent partially-observable MDP (Dec-POMDP) task.
The task consists of a tuple $G\textup{\texttt{=}}\langle I, S, A, P, R, \Omega, O, n, t, \gamma\rangle$, where $I$ is the finite set of $n+t$ agents interacting and communicating with each other. $n$ of the agents are benign, and $t$ of them are malicious whose communication channels (instead of actions) are attacked. In Dec-POMDP, $S,A,\Omega$ are the state, action and observation space. $O$ is the observation function, and $P$ is the transition function of the environment. $R$ is the global reward function, and $\gamma\in[0, 1)$ is the discount factor. The whole system is trained by Q-learning, where the goal is to collectively maximize the global return $\mathbb{E}_p[\sum_{t=0}^{\infty} \gamma^t R_t]$. 

\begin{figure}[!h]
\label{pipeline}
\centering
\includegraphics[width=0.9\linewidth]{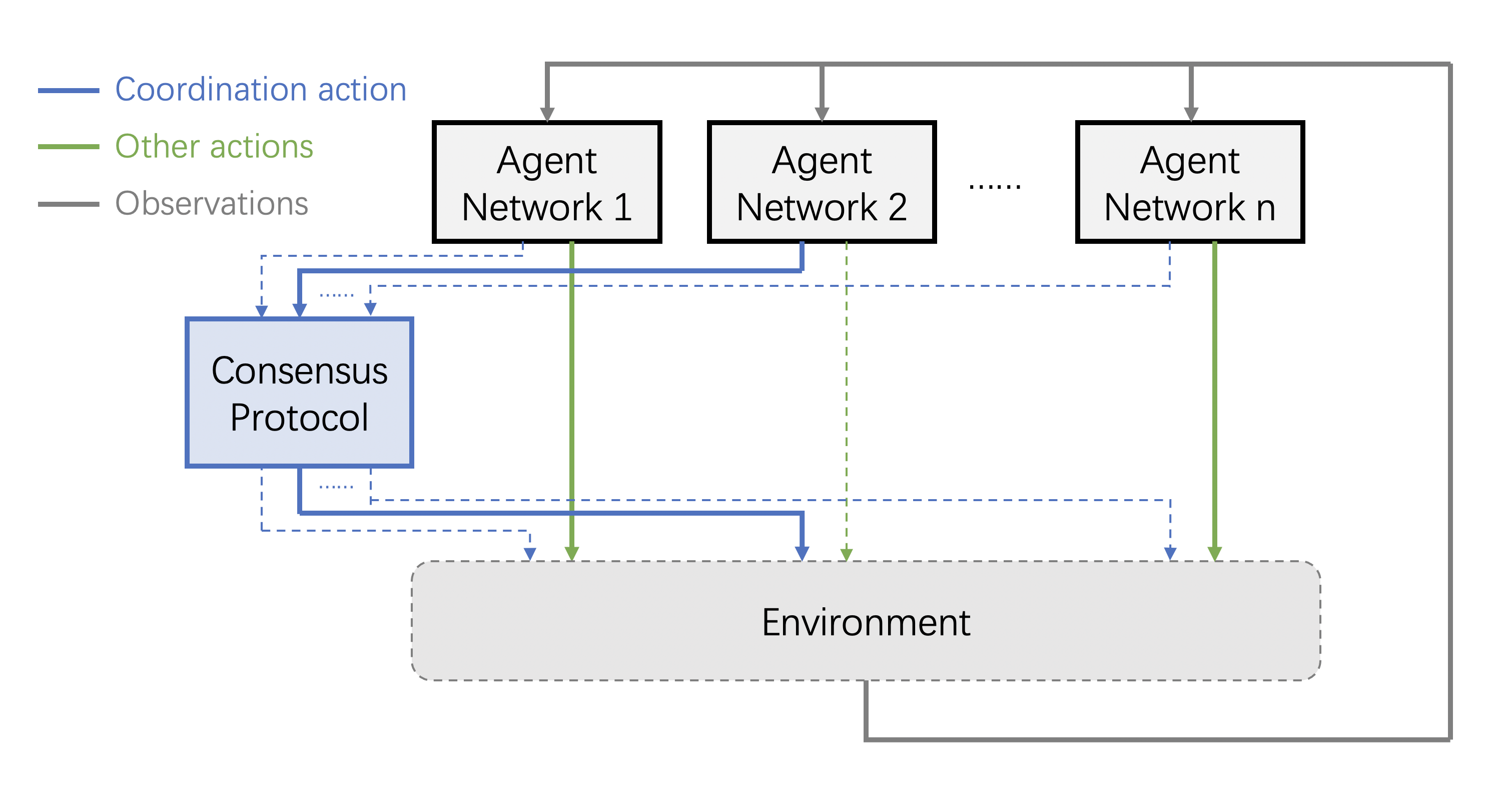}
\caption{The learnable action selection with the consensus protocol.}
\label{pipeline}
\end{figure}




The whole pipeline is shown in Figure \ref{pipeline}. The consensus process serves as a subprocess within the broader POMDP framework. At each time step, $s \in S$ is the state of the environment. Each agent $i$ receives a partial observation of the state $o_i$ drawn from $\Omega$, according to the observation function $O(s, i)$. Then the agent selects an action $a_i$ from $A$, based on its local action-observation history $\tau_i\in \mathrm{T}\equiv(\Omega\times A)^*\times \Omega$. During the action selection process, the agent decices whether to select a normal action or propose a coordination consensus on a certain target. 

For instance, consider a predator-prey game, $o_i$ encapsulates the states of neighboring grids relative to agent $i$; and $a_i$ includes possible movements like \textit{go up}, \textit{go down}, \textit{go left} and \textit{go right}, alongside the consensus-aware action \textit{propose to catch}. Within the consensus process, $M_i^{0}=1$ if $a_i$ is set to \textit{propose to catch}, indicating the intention to pursue the prey, as depicted in Figure \ref{pipeline} by an agent emitting both a dashed green line and a solid blue line. In contrast, $M_i^{0}=0$ under other actions, meaning that there's no propose, as depicted in Figure \ref{pipeline} by an agent emitting both a solid green line and a dashed blue line. An agent would select an $a_i$ that maximizes its local Q-value based on $o_i$. When the agent's observation indicates a potential for coordination (e.g., $o_i$ reveals proximity to the prey), the action \textit{propose to catch} is chosen by the policy network, and $M_i^{0}$ switches to $1$, participating in the consensus process. If agents reach consensus on \textit{catch} after the communication according to the protocol, the agent will then catch the prey. If agents reach consensus on not catching, the agents with \textit{catch} proposals just give up cooperative catching at this step and do nothing.

\begin{figure}
\centering
    \subfloat[4bane\_vs\_1hM]{\includegraphics[width=0.45\textwidth]{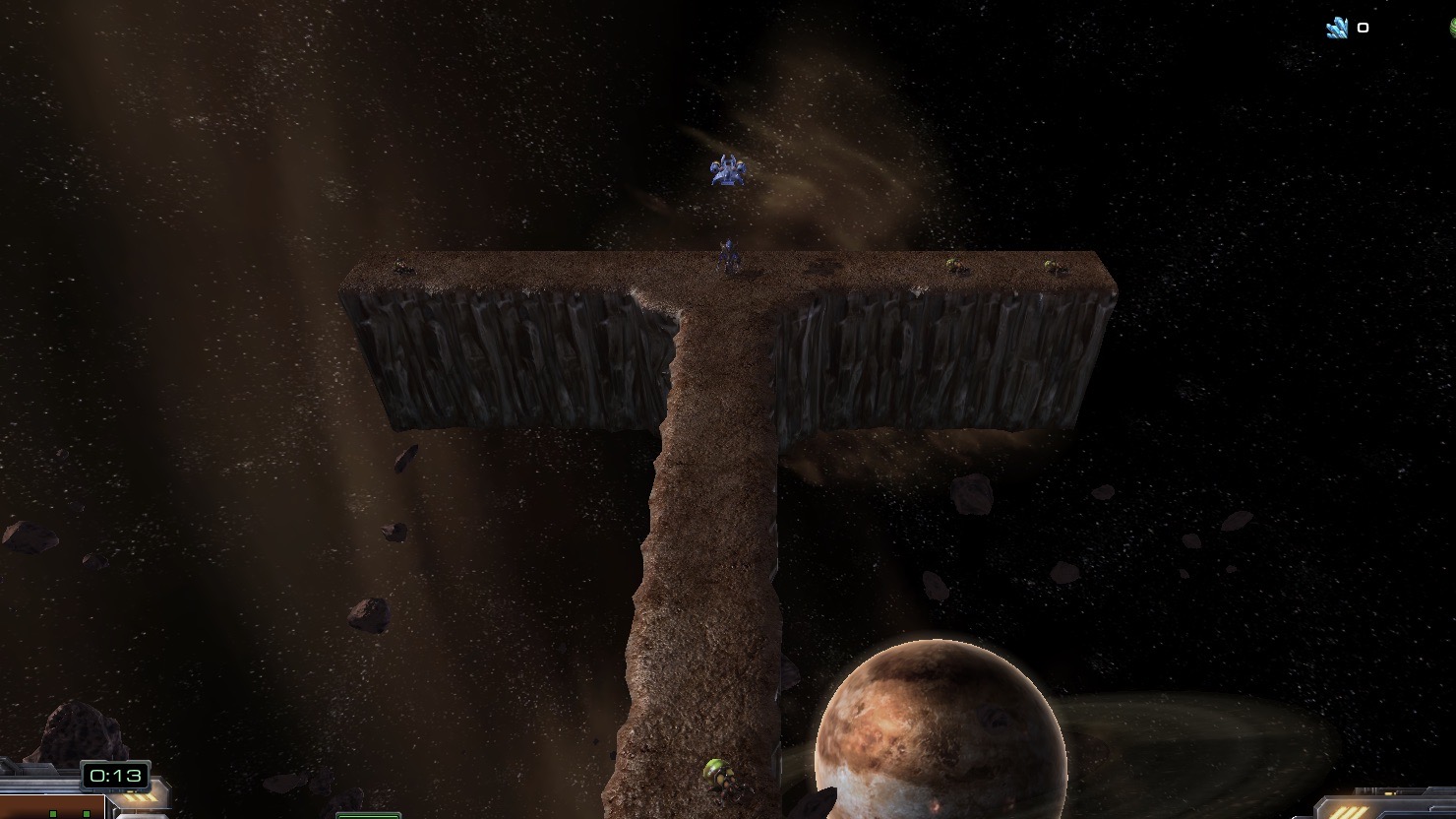}}\\
    \subfloat[3z\_vs\_1r]{\includegraphics[width=0.45\textwidth]{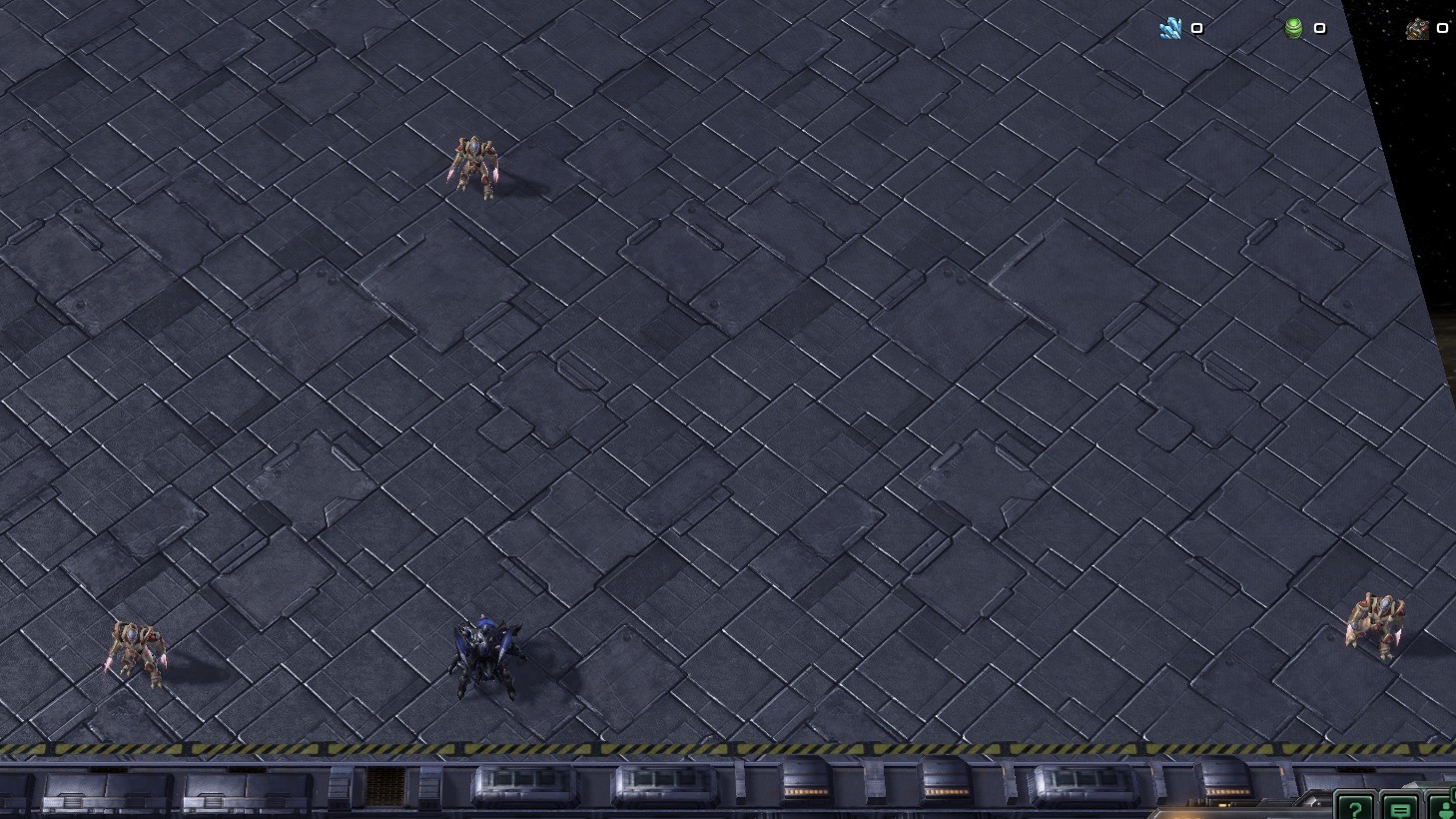}}
    \caption{The screenshot of 2 StarCraft II enironments.}
    \label{fig:sc2}
\end{figure}

\section{Experiments}
\label{exp}
While Theorem \ref{thm:protocol} ensures that the proposed protocol guarantees safe coordination in the Imperfect Byzantine Generals Problem (IBGP), it raises intriguing questions: (i) Is the IBGP protocol applicable to the MARL setting, as discussed in Section 4.2, and does it outperform baselines in terms of robustness? (ii) How can we utilize the protocol in real-world applications? In Section \ref{exp:main}, we compare the consensus-based method—integrating the IBGP protocol into the MARL pipeline—with existing RL-based approaches to evaluate robustness under attack. To address the second question, Section \ref{exp:sensor} presents a case study in which we apply a consensus-based algorithm to address the Sensor Network problem. Implementation details and hyperparameters are included in the Appendix \ref{apd:impl}.

\subsection{Main Results}
\label{exp:main}
The experiment includes four kinds of environments, denoted as Env$(n,m,k,t)$. Here, $n$ represents the number of benign agents, $m$ is the number of targets, $k$ is the coordination threshold, and $t$ is the number of attackers.
\begin{itemize}
    \item \textit{Predator-prey}: It is modified from the well-known predator-prey environment \cite{stone2000multiagent}, requiring several predators to hunt the prey together. We test single-target Predator-prey$(4,1,2,1)$, multi-target Predator-prey$(5,2,2,1)$, and large-scale Predator-prey$(20,4,2,2)$ and Predator-prey$(20,1,4,2)$.
    \item \textit{Hallway}: It is introduced in \cite{tonghan2020learning}, requiring several agents to reach the destination simultaneously. Besides the environment with the original parameters Hallway$(3,1,2,1)$, we also test its scalability with more agents and a larger coordination threshold in Hallway$(10,1,5,2)$.
    \item \textit{4bane\_vs\_1hM}: It is built on the SMAC benchmark (StarCraft Multi-agent Challenge) \cite{mikayel2019starcraft}. 4bane\_vs\_1hM is an environment modified from 3bane\_vs\_1hM, which is used in previous communicative MARL research. We add an agent because of the additional communicative attack. It requires the agents (3 banelings) to attack the enemy simultaneously; otherwise, the enemy will heal itself and no longer can be killed. 
    \item \textit{3z\_vs\_1r}: We also create a new SC2 environment, in which two agents (zealots) are required to attack the enemy (a roach) simultaneously. If they fail to do so, the enemy's long-distance firepower will successively eliminate the agents, resulting in insufficient damage to defeat the enemy. This task requires both resilience against attacks and efficiency. Waiting for coordinated action puts the agents in danger of being killed by the enemy. Therefore, it is essential to find a balance between coordination with the fewest agents possible and defense against attacks. This task serves as a challenging benchmark and highlights the potential for improving efficiency and robustness simultaneously.
\end{itemize}

\begin{table*}
    \centering
    \begin{tabular}{c|c|c|c|c}
        \toprule
        &IBGP Protocol&Recursive training&AME&ADMAC\\
        \midrule
         Predator-prey$(4,1,2,1)$&$96.1\pm5.4\%$&$0\%$&$62.4\pm20.4\%$&$25.6\pm13.2\%$\\
        \hline
         Predator-prey$(5,2,2,1)$&$97.9\pm2.9\%$&$3.7\pm3.5\%$&$42.6\pm13.0\%$&$14.0\pm7.7\%$\\
        \hline
         Predator-prey$(20,4,2,2)$&$100.0\pm0\%$&$0\%$&$79.3\pm12.1\%$&/\\
        \hline
         Predator-prey$(20,1,4,2)$&$100.0\pm0\%$&$16.5\pm16.4\%$&$100.0\pm0\%$&$54.1\pm20.3\%$\\
        \hline
        Hallway$(3,1,2,1)$&$96.7\pm4.7\%$&$0\%$&$6.3\pm6.3\%$&$6.4\pm4.8\%$\\
        \hline
         Hallway$(10,1,5,2)$&$100.0\pm0\%$&/&$13.1\pm7.2\%$&$7.4\pm10.4\%$\\
        \hline
         4bane\_vs\_1hM$(4,1,3,1)$&$98.4\pm2.2\%$&$20.4\pm6.4\%$&$92.9\pm4.7\%$&$64.5\pm13.6\%$\\
        \hline
         3z\_vs\_1r$(3,1,2,1)$&$51.5\pm2.6\%$&$6.2\pm0.8\%$&/&/\\
        \bottomrule
    \end{tabular}
    \caption{The table illustrates the robustness percentages and their standard deviation of different environments and algorithms. A higher robustness percentage indicates a higher level of resilience against zero-shot communicative attacks. Algorithms that fail to converge during training are marked with '/'.}
    \label{tab:robust}
\end{table*}

We conduct experiments to demonstrate the zero-shot robustness of the IBGP protocol in the environments. Although the IBGP protocol defines how to communicate, the agents still need to learn a well-performing policy, which includes decisions on how to move and when to propose a consensus. To evaluate the zero-shot robustness of algorithms, we define the robustness percentage as the ratio of performance with attackers (during the testing phase) to performance with no attackers (during the training phase). Specifically, we train attackers' communication to harm the benign agents' performance to the greatest extent in the testing phase, thus displaying zero-shot robustness against any attack scheme. If the algorithm is not robust against communicative attacks, the result of the testing phase will significantly decay.

Table \ref{tab:robust} indicates the robustness percentage of the IBGP protocol in the first column, while the second column displays the ratio of recursive training, which means that the training process of agents and attackers is repeated. Recursive training is one of the contributions of the algorithm in \cite{xue2022mis}. The last two column provides a comparison with AME \cite{sun2023certifiably} and ADMAC \cite{Yu_Qiu_Yao_Shen_Zhang_Wang_2024}.

Analysis of Table \ref{tab:robust} reveals that the IBGP Protocol maintains its performance from training to testing phases. However, recursive training is not consistently robust when transitioning to testing in some environments. This is reasonable since the baseline paper \cite{xue2022mis} prioritizes adaption to attackers rather than zero-shot robustness. The AME algorithm performs well in some environments during testing, although it experiences degradation in performance in others. Since the AME algorithm is designed to be robust through its novel design of taking the majority, we think that the performance decay is due to the fact that the action in the majority may be multiple, making attacks still possible. The ADMAC algorithm also exhibits a low level of robustness in our environments, likely because its reliability estimator was trained on specific types of attacks, with only minor disturbances evaluated in the original paper. However, in this work, we focus on robustness against arbitrary attacks, which leads to ADMAC underperforming. In summary, policy learning with the IBGP Protocol is effective in fostering robust coordination against zero-shot communicative attacks.




\begin{figure}
    \centering
    \includegraphics[width=0.9\linewidth]{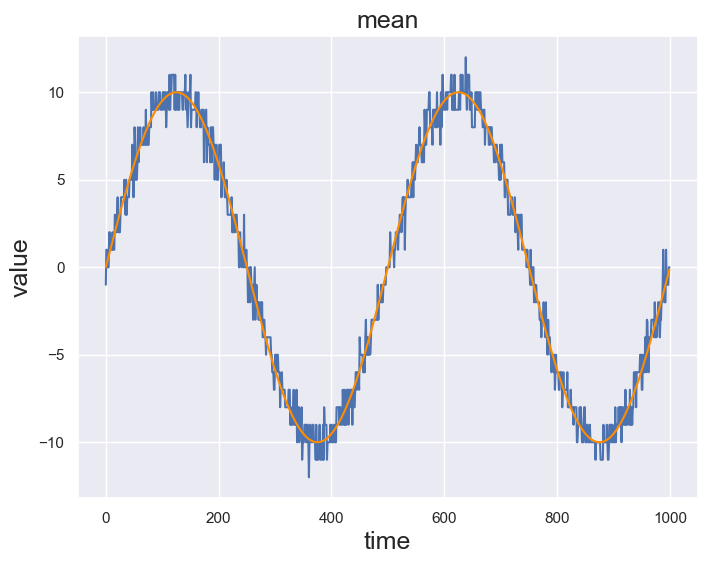}
    \caption{The simulation result of the Sensor Networks Problem.}
    \label{fig:sensor}
\end{figure}

\begin{figure*}
    \centering
    \includegraphics[width=\textwidth]{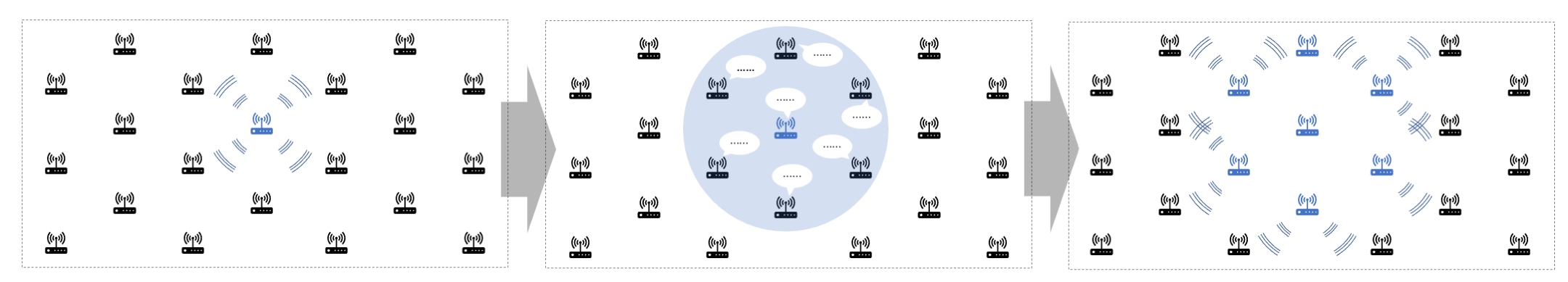}
    \caption{The pipeline of applying a consensus-based algorithm on the sensor network problem.}
    \label{fig:sensor_graph}
\end{figure*}

\subsection{Application in Sensor Networks}
\label{exp:sensor}

In this paper, we address the sensor network problem with communicative attacks, building on the continuous sensor network problem discussed in \cite{Olfati2005Consensus}. We begin by defining the state (position) of the target as $s(t)$, where $t$ is the time. For simplicity, we assume a one-dimensional state $s(t)\in\mathbb{R}$, as multi-dimensional cases are extensions of this. The sensor network consists of $n$ sensors located at $pos_1,pos_2,\cdots,pos_n$, with sensor $i$ observing $o_i(t)=s(t)+\epsilon$ if it is close to the target ($dist(s(t),pos_i)<\eta$). $\epsilon$ is the noise function. The agents' goal is to safely share the target position among the sensor network, while the attackers aim to break the system's robustness. We denote the agent's belief on the state (target position) as $a_{i}(t)$, and the consistency of the whole system as $a_1(t)=a_2(t)=\cdots=a_n(t)$.

\paragraph{Consensus-based algorithm} We propose a decentralized algorithm based on the IBGP protocol for the sensor network problem. Each sensor is activated if it observes the target or is involved in the protocol and reaches a new consensus on the target position. Since the observed target position includes a noise term, we first discretize the observed position to avoid observation disagreement with other sensors. When a sensor is activated, it notifies its neighborhood and proposes an IBGP protocol on the new target position. In the consensus, the agreement on the scope and value of the protocol is guaranteed by BGP-related protocols, which is not the emphasis of this part. By recursively accomplishing the IBGP protocols until there are no more activating sensors, the sharing process is completed, and the entire sensor network agrees on the target position. The three recursive steps are listed below, and Figure \ref{fig:sensor_graph} shows the three steps.
\begin{enumerate}
    \item First step: In each timestep $t$, agent $i$ observes a new observation $o_i(t)=s(t)+\epsilon$, where $s_t$ is the state (the true position of the target), and $\epsilon$ is a noise term. The agent then discretizes this signal to form a new belief $a_{i,t}=\mathrm{discretize}(o_i(t))$. If the new signal updates the agent's belief on the target position, it proceeds to the second step.
    \item Second step: The agent proposes a neighborhood consensus using the $(1,1)$-protocol.
    \item Third step: If other agents in the consensus update their belief, the agent proposes a consensus in its own neighborhood. This process is operated iteratively until there are no new consensus proposals.
\end{enumerate}

In Figure \ref{fig:sensor}, the simulation shows that the belief of all agents is consistent with the true signal. As the target position changes, the believed position moves along with it, indicating that consensus on the target position is maintained throughout. More details can be found in Appendix \ref{apd:sensor}.


\section{Conclusion}
In this study, we introduce the Imperfect Byzantine Generals Problem (IBGP) as a framework for multi-agent coordination in the presence of compromised agents. The consensus protocol we have developed for IBGP effectively ensures zero-shot robustness. While our analysis and experiments demonstrates its potential applicability to practical tasks, the proposed IBGP protocol is only effective in environments where a specific target exists, rather than arbitrary RL environments. In future research, we intend to address this limitation by exploring the adaptability of a generalized trainable framework for consensus protocols to more complex and realistic settings.




\bibliography{neurips_2024}


\begin{thebibliography}{26}


\ifx \showCODEN    \undefined \def \showCODEN     #1{\unskip}     \fi
\ifx \showDOI      \undefined \def \showDOI       #1{#1}\fi
\ifx \showISBNx    \undefined \def \showISBNx     #1{\unskip}     \fi
\ifx \showISBNxiii \undefined \def \showISBNxiii  #1{\unskip}     \fi
\ifx \showISSN     \undefined \def \showISSN      #1{\unskip}     \fi
\ifx \showLCCN     \undefined \def \showLCCN      #1{\unskip}     \fi
\ifx \shownote     \undefined \def \shownote      #1{#1}          \fi
\ifx \showarticletitle \undefined \def \showarticletitle #1{#1}   \fi
\ifx \showURL      \undefined \def \showURL       {\relax}        \fi
\providecommand\bibfield[2]{#2}
\providecommand\bibinfo[2]{#2}
\providecommand\natexlab[1]{#1}
\providecommand\showeprint[2][]{arXiv:#2}

\bibitem[\protect\citeauthoryear{B{\"o}hmer, Kurin, and Whiteson}{B{\"o}hmer et~al\mbox{.}}{2020}]%
        {bohmer2020deep}
\bibfield{author}{\bibinfo{person}{Wendelin B{\"o}hmer}, \bibinfo{person}{Vitaly Kurin}, {and} \bibinfo{person}{Shimon Whiteson}.} \bibinfo{year}{2020}\natexlab{}.
\newblock \showarticletitle{Deep coordination graphs}. In \bibinfo{booktitle}{\emph{Proceedings of the 37th International Conference on Machine Learning}}.
\newblock


\bibitem[\protect\citeauthoryear{Castro and Liskov}{Castro and Liskov}{1999}]%
        {castro1999practical}
\bibfield{author}{\bibinfo{person}{Miguel Castro} {and} \bibinfo{person}{Barbara Liskov}.} \bibinfo{year}{1999}\natexlab{}.
\newblock \showarticletitle{Practical Byzantine Fault Tolerance}. In \bibinfo{booktitle}{\emph{Proceedings of the Third Symposium on Operating Systems Design and Implementation}} (New Orleans, Louisiana, USA) \emph{(\bibinfo{series}{OSDI '99})}. \bibinfo{publisher}{USENIX Association}, \bibinfo{address}{USA}, \bibinfo{pages}{173–186}.
\newblock
\showISBNx{1880446391}


\bibitem[\protect\citeauthoryear{ChonghanYu;}{ChonghanYu;}{2024}]%
        {Wu_Huang_Yu_2024}
\bibfield{author}{\bibinfo{person}{XiaoranWu;~ZienHuang; ChonghanYu;}.} \bibinfo{year}{2024}\natexlab{}.
\newblock \showarticletitle{Animating the Past: Reconstruct Trilobite via Video Generation}.
\newblock  (\bibinfo{year}{2024}).
\newblock
\urldef\tempurl%
\url{https://doi.org/10.12074/202410.00084}
\showDOI{\tempurl}


\bibitem[\protect\citeauthoryear{Dorri, Kanhere, and Jurdak}{Dorri et~al\mbox{.}}{2018}]%
        {dorri2018multi}
\bibfield{author}{\bibinfo{person}{Ali Dorri}, \bibinfo{person}{Salil~S Kanhere}, {and} \bibinfo{person}{Raja Jurdak}.} \bibinfo{year}{2018}\natexlab{}.
\newblock \showarticletitle{Multi-agent systems: A survey}.
\newblock \bibinfo{journal}{\emph{Ieee Access}}  \bibinfo{volume}{6} (\bibinfo{year}{2018}), \bibinfo{pages}{28573--28593}.
\newblock


\bibitem[\protect\citeauthoryear{Fan, Zhang, and Wang}{Fan et~al\mbox{.}}{2014}]%
        {fan2014bipartite}
\bibfield{author}{\bibinfo{person}{Ming-Can Fan}, \bibinfo{person}{Hai-Tao Zhang}, {and} \bibinfo{person}{Miaomiao Wang}.} \bibinfo{year}{2014}\natexlab{}.
\newblock \showarticletitle{Bipartite flocking for multi-agent systems}.
\newblock \bibinfo{journal}{\emph{Communications in Nonlinear Science and Numerical Simulation}} \bibinfo{volume}{19}, \bibinfo{number}{9} (\bibinfo{year}{2014}), \bibinfo{pages}{3313--3322}.
\newblock


\bibitem[\protect\citeauthoryear{Fu and Wang}{Fu and Wang}{2014}]%
        {fu2014adaptive}
\bibfield{author}{\bibinfo{person}{Junjie Fu} {and} \bibinfo{person}{Jinzhi Wang}.} \bibinfo{year}{2014}\natexlab{}.
\newblock \showarticletitle{Adaptive coordinated tracking of multi-agent systems with quantized information}.
\newblock \bibinfo{journal}{\emph{Systems \& Control Letters}}  \bibinfo{volume}{74} (\bibinfo{year}{2014}), \bibinfo{pages}{115--125}.
\newblock


\bibitem[\protect\citeauthoryear{Hong, Zheng, Chen, Cheng, Wang, Zhang, Wang, Yau, Lin, Zhou, et~al\mbox{.}}{Hong et~al\mbox{.}}{2023}]%
        {hong2023metagpt}
\bibfield{author}{\bibinfo{person}{Sirui Hong}, \bibinfo{person}{Xiawu Zheng}, \bibinfo{person}{Jonathan Chen}, \bibinfo{person}{Yuheng Cheng}, \bibinfo{person}{Jinlin Wang}, \bibinfo{person}{Ceyao Zhang}, \bibinfo{person}{Zili Wang}, \bibinfo{person}{Steven Ka~Shing Yau}, \bibinfo{person}{Zijuan Lin}, \bibinfo{person}{Liyang Zhou}, {et~al\mbox{.}}} \bibinfo{year}{2023}\natexlab{}.
\newblock \showarticletitle{Metagpt: Meta programming for multi-agent collaborative framework}.
\newblock \bibinfo{journal}{\emph{arXiv preprint arXiv:2308.00352}} (\bibinfo{year}{2023}).
\newblock


\bibitem[\protect\citeauthoryear{Kang, Wang, and de~Melo}{Kang et~al\mbox{.}}{2020}]%
        {kang2020incorporating}
\bibfield{author}{\bibinfo{person}{Yipeng Kang}, \bibinfo{person}{Tonghan Wang}, {and} \bibinfo{person}{Gerard de Melo}.} \bibinfo{year}{2020}\natexlab{}.
\newblock \showarticletitle{Incorporating pragmatic reasoning communication into emergent language}.
\newblock \bibinfo{journal}{\emph{Advances in Neural Information Processing Systems}}  \bibinfo{volume}{33} (\bibinfo{year}{2020}), \bibinfo{pages}{10348--10359}.
\newblock


\bibitem[\protect\citeauthoryear{Kang, Wang, Yang, Wu, and Zhang}{Kang et~al\mbox{.}}{2022}]%
        {kang2022non}
\bibfield{author}{\bibinfo{person}{Yipeng Kang}, \bibinfo{person}{Tonghan Wang}, \bibinfo{person}{Qianlan Yang}, \bibinfo{person}{Xiaoran Wu}, {and} \bibinfo{person}{Chongjie Zhang}.} \bibinfo{year}{2022}\natexlab{}.
\newblock \showarticletitle{Non-Linear Coordination Graphs}.
\newblock \bibinfo{journal}{\emph{Advances in Neural Information Processing Systems}}  \bibinfo{volume}{35} (\bibinfo{year}{2022}), \bibinfo{pages}{25655--25666}.
\newblock


\bibitem[\protect\citeauthoryear{Kishishita and Ozaki}{Kishishita and Ozaki}{2020}]%
        {daiki2020public}
\bibfield{author}{\bibinfo{person}{Daiki Kishishita} {and} \bibinfo{person}{Hiroyuki Ozaki}.} \bibinfo{year}{2020}\natexlab{}.
\newblock \showarticletitle{Public goods game with ambiguous threshold}.
\newblock \bibinfo{journal}{\emph{Economics Letters}}  \bibinfo{volume}{191} (\bibinfo{year}{2020}), \bibinfo{pages}{109165}.
\newblock
\showISSN{0165-1765}
\urldef\tempurl%
\url{https://doi.org/10.1016/j.econlet.2020.109165}
\showDOI{\tempurl}


\bibitem[\protect\citeauthoryear{Lamport, Shostak, and Pease}{Lamport et~al\mbox{.}}{1982}]%
        {lamport1982byzantine}
\bibfield{author}{\bibinfo{person}{Leslie Lamport}, \bibinfo{person}{Robert Shostak}, {and} \bibinfo{person}{Marshall Pease}.} \bibinfo{year}{1982}\natexlab{}.
\newblock \showarticletitle{The Byzantine Generals Problem}.
\newblock \bibinfo{journal}{\emph{ACM Trans. Program. Lang. Syst.}} \bibinfo{volume}{4}, \bibinfo{number}{3} (\bibinfo{date}{jul} \bibinfo{year}{1982}), \bibinfo{pages}{382–401}.
\newblock
\showISSN{0164-0925}
\urldef\tempurl%
\url{https://doi.org/10.1145/357172.357176}
\showDOI{\tempurl}


\bibitem[\protect\citeauthoryear{Liu, Xie, and Zhang}{Liu et~al\mbox{.}}{2014}]%
        {liu2014containment}
\bibfield{author}{\bibinfo{person}{Shuai Liu}, \bibinfo{person}{Lihua Xie}, {and} \bibinfo{person}{Huanshui Zhang}.} \bibinfo{year}{2014}\natexlab{}.
\newblock \showarticletitle{Containment control of multi-agent systems by exploiting the control inputs of neighbors}.
\newblock \bibinfo{journal}{\emph{International Journal of Robust and Nonlinear Control}} \bibinfo{volume}{24}, \bibinfo{number}{17} (\bibinfo{year}{2014}), \bibinfo{pages}{2803--2818}.
\newblock


\bibitem[\protect\citeauthoryear{Olfati-Saber}{Olfati-Saber}{2006}]%
        {olfati2006flocking}
\bibfield{author}{\bibinfo{person}{Reza Olfati-Saber}.} \bibinfo{year}{2006}\natexlab{}.
\newblock \showarticletitle{Flocking for multi-agent dynamic systems: Algorithms and theory}.
\newblock \bibinfo{journal}{\emph{IEEE Transactions on automatic control}} \bibinfo{volume}{51}, \bibinfo{number}{3} (\bibinfo{year}{2006}), \bibinfo{pages}{401--420}.
\newblock


\bibitem[\protect\citeauthoryear{Olfati-Saber and Murray}{Olfati-Saber and Murray}{2004}]%
        {olfati2004consensus}
\bibfield{author}{\bibinfo{person}{Reza Olfati-Saber} {and} \bibinfo{person}{Richard~M Murray}.} \bibinfo{year}{2004}\natexlab{}.
\newblock \showarticletitle{Consensus problems in networks of agents with switching topology and time-delays}.
\newblock \bibinfo{journal}{\emph{IEEE Transactions on automatic control}} \bibinfo{volume}{49}, \bibinfo{number}{9} (\bibinfo{year}{2004}), \bibinfo{pages}{1520--1533}.
\newblock


\bibitem[\protect\citeauthoryear{Olfati-Saber and Shamma}{Olfati-Saber and Shamma}{2005}]%
        {Olfati2005Consensus}
\bibfield{author}{\bibinfo{person}{R. Olfati-Saber} {and} \bibinfo{person}{J.S. Shamma}.} \bibinfo{year}{2005}\natexlab{}.
\newblock \showarticletitle{Consensus Filters for Sensor Networks and Distributed Sensor Fusion}. In \bibinfo{booktitle}{\emph{Proceedings of the 44th IEEE Conference on Decision and Control}}. \bibinfo{pages}{6698--6703}.
\newblock
\urldef\tempurl%
\url{https://doi.org/10.1109/CDC.2005.1583238}
\showDOI{\tempurl}


\bibitem[\protect\citeauthoryear{Rabin}{Rabin}{1983}]%
        {rabin1983randomized}
\bibfield{author}{\bibinfo{person}{Michael~O. Rabin}.} \bibinfo{year}{1983}\natexlab{}.
\newblock \showarticletitle{Randomized byzantine generals}. In \bibinfo{booktitle}{\emph{24th Annual Symposium on Foundations of Computer Science (sfcs 1983)}}. \bibinfo{pages}{403--409}.
\newblock
\urldef\tempurl%
\url{https://doi.org/10.1109/SFCS.1983.48}
\showDOI{\tempurl}


\bibitem[\protect\citeauthoryear{Samvelyan, Rashid, de~Witt, Farquhar, Nardelli, Rudner, Hung, Torr, Foerster, and Whiteson}{Samvelyan et~al\mbox{.}}{2019}]%
        {mikayel2019starcraft}
\bibfield{author}{\bibinfo{person}{Mikayel Samvelyan}, \bibinfo{person}{Tabish Rashid}, \bibinfo{person}{Christian~Schr{\"{o}}der de Witt}, \bibinfo{person}{Gregory Farquhar}, \bibinfo{person}{Nantas Nardelli}, \bibinfo{person}{Tim G.~J. Rudner}, \bibinfo{person}{Chia{-}Man Hung}, \bibinfo{person}{Philip H.~S. Torr}, \bibinfo{person}{Jakob~N. Foerster}, {and} \bibinfo{person}{Shimon Whiteson}.} \bibinfo{year}{2019}\natexlab{}.
\newblock \showarticletitle{The StarCraft Multi-Agent Challenge}.
\newblock \bibinfo{journal}{\emph{CoRR}}  \bibinfo{volume}{abs/1902.04043} (\bibinfo{year}{2019}).
\newblock
\showeprint[arXiv]{1902.04043}
\urldef\tempurl%
\url{http://arxiv.org/abs/1902.04043}
\showURL{%
\tempurl}


\bibitem[\protect\citeauthoryear{Son, Kim, Kang, Hostallero, and Yi}{Son et~al\mbox{.}}{2019}]%
        {son2019qtran}
\bibfield{author}{\bibinfo{person}{Kyunghwan Son}, \bibinfo{person}{Daewoo Kim}, \bibinfo{person}{Wan~Ju Kang}, \bibinfo{person}{David~Earl Hostallero}, {and} \bibinfo{person}{Yung Yi}.} \bibinfo{year}{2019}\natexlab{}.
\newblock \showarticletitle{{QTRAN}: Learning to Factorize with Transformation for Cooperative Multi-Agent Reinforcement Learning}. In \bibinfo{booktitle}{\emph{Proceedings of the 36th International Conference on Machine Learning}} \emph{(\bibinfo{series}{Proceedings of Machine Learning Research}, Vol.~\bibinfo{volume}{97})}, \bibfield{editor}{\bibinfo{person}{Kamalika Chaudhuri} {and} \bibinfo{person}{Ruslan Salakhutdinov}} (Eds.). \bibinfo{publisher}{PMLR}, \bibinfo{pages}{5887--5896}.
\newblock
\urldef\tempurl%
\url{https://proceedings.mlr.press/v97/son19a.html}
\showURL{%
\tempurl}


\bibitem[\protect\citeauthoryear{Stone and Veloso}{Stone and Veloso}{2000}]%
        {stone2000multiagent}
\bibfield{author}{\bibinfo{person}{Peter Stone} {and} \bibinfo{person}{Manuela Veloso}.} \bibinfo{year}{2000}\natexlab{}.
\newblock \showarticletitle{Multiagent systems: A survey from a machine learning perspective}.
\newblock \bibinfo{journal}{\emph{Autonomous Robots}}  \bibinfo{volume}{8} (\bibinfo{year}{2000}), \bibinfo{pages}{345--383}.
\newblock


\bibitem[\protect\citeauthoryear{Sun, Zheng, Hassanzadeh, Liang, Feizi, Ganesh, and Huang}{Sun et~al\mbox{.}}{2023}]%
        {sun2023certifiably}
\bibfield{author}{\bibinfo{person}{Yanchao Sun}, \bibinfo{person}{Ruijie Zheng}, \bibinfo{person}{Parisa Hassanzadeh}, \bibinfo{person}{Yongyuan Liang}, \bibinfo{person}{Soheil Feizi}, \bibinfo{person}{Sumitra Ganesh}, {and} \bibinfo{person}{Furong Huang}.} \bibinfo{year}{2023}\natexlab{}.
\newblock \showarticletitle{Certifiably Robust Policy Learning against Adversarial Multi-Agent Communication}. In \bibinfo{booktitle}{\emph{The Eleventh International Conference on Learning Representations}}.
\newblock
\urldef\tempurl%
\url{https://openreview.net/forum?id=dCOL0inGl3e}
\showURL{%
\tempurl}


\bibitem[\protect\citeauthoryear{Wang*, Wang*, Zheng, and Zhang}{Wang* et~al\mbox{.}}{2020}]%
        {tonghan2020learning}
\bibfield{author}{\bibinfo{person}{Tonghan Wang*}, \bibinfo{person}{Jianhao Wang*}, \bibinfo{person}{Chongyi Zheng}, {and} \bibinfo{person}{Chongjie Zhang}.} \bibinfo{year}{2020}\natexlab{}.
\newblock \showarticletitle{Learning Nearly Decomposable Value Functions Via Communication Minimization}. In \bibinfo{booktitle}{\emph{International Conference on Learning Representations}}.
\newblock
\urldef\tempurl%
\url{https://openreview.net/forum?id=HJx-3grYDB}
\showURL{%
\tempurl}


\bibitem[\protect\citeauthoryear{Wu, Bansal, Zhang, Wu, Zhang, Zhu, Li, Jiang, Zhang, and Wang}{Wu et~al\mbox{.}}{2023}]%
        {wu2023autogen}
\bibfield{author}{\bibinfo{person}{Qingyun Wu}, \bibinfo{person}{Gagan Bansal}, \bibinfo{person}{Jieyu Zhang}, \bibinfo{person}{Yiran Wu}, \bibinfo{person}{Shaokun Zhang}, \bibinfo{person}{Erkang Zhu}, \bibinfo{person}{Beibin Li}, \bibinfo{person}{Li Jiang}, \bibinfo{person}{Xiaoyun Zhang}, {and} \bibinfo{person}{Chi Wang}.} \bibinfo{year}{2023}\natexlab{}.
\newblock \showarticletitle{Autogen: Enabling next-gen llm applications via multi-agent conversation framework}.
\newblock \bibinfo{journal}{\emph{arXiv preprint arXiv:2308.08155}} (\bibinfo{year}{2023}).
\newblock


\bibitem[\protect\citeauthoryear{Xue, Qiu, An, Rabinovich, Obraztsova, and Yeo}{Xue et~al\mbox{.}}{2022}]%
        {xue2022mis}
\bibfield{author}{\bibinfo{person}{Wanqi Xue}, \bibinfo{person}{Wei Qiu}, \bibinfo{person}{Bo An}, \bibinfo{person}{Zinovi Rabinovich}, \bibinfo{person}{Svetlana Obraztsova}, {and} \bibinfo{person}{Chai~Kiat Yeo}.} \bibinfo{year}{2022}\natexlab{}.
\newblock \showarticletitle{Mis-Spoke or Mis-Lead: Achieving Robustness in Multi-Agent Communicative Reinforcement Learning} \emph{(\bibinfo{series}{AAMAS '22})}. \bibinfo{publisher}{International Foundation for Autonomous Agents and Multiagent Systems}, \bibinfo{address}{Richland, SC}.
\newblock
\showISBNx{9781450392136}


\bibitem[\protect\citeauthoryear{Yu, Qiu, Yao, Shen, Zhang, and Wang}{Yu et~al\mbox{.}}{2024}]%
        {Yu_Qiu_Yao_Shen_Zhang_Wang_2024}
\bibfield{author}{\bibinfo{person}{Lebin Yu}, \bibinfo{person}{Yunbo Qiu}, \bibinfo{person}{Quanming Yao}, \bibinfo{person}{Yuan Shen}, \bibinfo{person}{Xudong Zhang}, {and} \bibinfo{person}{Jian Wang}.} \bibinfo{year}{2024}\natexlab{}.
\newblock \showarticletitle{Robust Communicative Multi-Agent Reinforcement Learning with Active Defense}.
\newblock \bibinfo{journal}{\emph{Proceedings of the AAAI Conference on Artificial Intelligence}} \bibinfo{volume}{38}, \bibinfo{number}{16} (\bibinfo{date}{Mar.} \bibinfo{year}{2024}), \bibinfo{pages}{17575--17582}.
\newblock
\urldef\tempurl%
\url{https://doi.org/10.1609/aaai.v38i16.29708}
\showDOI{\tempurl}


\bibitem[\protect\citeauthoryear{Yu, Chen, Cao, and Kurths}{Yu et~al\mbox{.}}{2009}]%
        {yu2009second}
\bibfield{author}{\bibinfo{person}{Wenwu Yu}, \bibinfo{person}{Guanrong Chen}, \bibinfo{person}{Ming Cao}, {and} \bibinfo{person}{J{\"u}rgen Kurths}.} \bibinfo{year}{2009}\natexlab{}.
\newblock \showarticletitle{Second-order consensus for multiagent systems with directed topologies and nonlinear dynamics}.
\newblock \bibinfo{journal}{\emph{IEEE Transactions on Systems, Man, and Cybernetics, Part B (Cybernetics)}} \bibinfo{volume}{40}, \bibinfo{number}{3} (\bibinfo{year}{2009}), \bibinfo{pages}{881--891}.
\newblock


\bibitem[\protect\citeauthoryear{Zhou, Luo, Villella, Yang, Rusu, Miao, Zhang, Alban, FADAKAR, Chen, Huang, Wen, Hassanzadeh, Graves, Zhu, Ni, Nguyen, Elsayed, Ammar, Cowen-Rivers, Ahilan, Tian, Palenicek, Rezaee, Yadmellat, Shao, chen, Zhang, Zhang, Hao, Liu, and Wang}{Zhou et~al\mbox{.}}{2021}]%
        {kober2021smarts}
\bibfield{author}{\bibinfo{person}{Ming Zhou}, \bibinfo{person}{Jun Luo}, \bibinfo{person}{Julian Villella}, \bibinfo{person}{Yaodong Yang}, \bibinfo{person}{David Rusu}, \bibinfo{person}{Jiayu Miao}, \bibinfo{person}{Weinan Zhang}, \bibinfo{person}{Montgomery Alban}, \bibinfo{person}{IMAN FADAKAR}, \bibinfo{person}{Zheng Chen}, \bibinfo{person}{Chongxi Huang}, \bibinfo{person}{Ying Wen}, \bibinfo{person}{Kimia Hassanzadeh}, \bibinfo{person}{Daniel Graves}, \bibinfo{person}{Zhengbang Zhu}, \bibinfo{person}{Yihan Ni}, \bibinfo{person}{Nhat Nguyen}, \bibinfo{person}{Mohamed Elsayed}, \bibinfo{person}{Haitham Ammar}, \bibinfo{person}{Alexander Cowen-Rivers}, \bibinfo{person}{Sanjeevan Ahilan}, \bibinfo{person}{Zheng Tian}, \bibinfo{person}{Daniel Palenicek}, \bibinfo{person}{Kasra Rezaee}, \bibinfo{person}{Peyman Yadmellat}, \bibinfo{person}{Kun Shao}, \bibinfo{person}{dong chen}, \bibinfo{person}{Baokuan Zhang}, \bibinfo{person}{Hongbo Zhang}, \bibinfo{person}{Jianye Hao}, \bibinfo{person}{Wulong Liu}, {and}
  \bibinfo{person}{Jun Wang}.} \bibinfo{year}{2021}\natexlab{}.
\newblock \showarticletitle{SMARTS: An Open-Source Scalable Multi-Agent RL Training School for Autonomous Driving}. In \bibinfo{booktitle}{\emph{Proceedings of the 2020 Conference on Robot Learning}} \emph{(\bibinfo{series}{Proceedings of Machine Learning Research}, Vol.~\bibinfo{volume}{155})}, \bibfield{editor}{\bibinfo{person}{Jens Kober}, \bibinfo{person}{Fabio Ramos}, {and} \bibinfo{person}{Claire Tomlin}} (Eds.). \bibinfo{publisher}{PMLR}, \bibinfo{pages}{264--285}.
\newblock
\urldef\tempurl%
\url{https://proceedings.mlr.press/v155/zhou21a.html}
\showURL{%
\tempurl}


\end{thebibliography}
\bibliographystyle{ACM-Reference-Format}


\newpage
\appendix

\section{Appendix / supplemental material}
\subsection{Diagram}
\label{diagram}
We provide the diagram about the results under all kinds of attacking strategies in Table \ref{tab:diagram}, given $n=5,t=1,k=3$. It illustrates that, no matter what the attack is, the IBGP protocol guarantees robustness with high probability $1-\max_{r}\{p(r_{tot}=r)\}$
\begin{table*}[]
    \centering
    \subfloat[Attacking strategies that directly causes the number of active agents $<k$.]{\begin{tabular}{|c|c|}
        \hline
        If $r_{tot}=1$ & \includegraphics[width=0.35\linewidth]{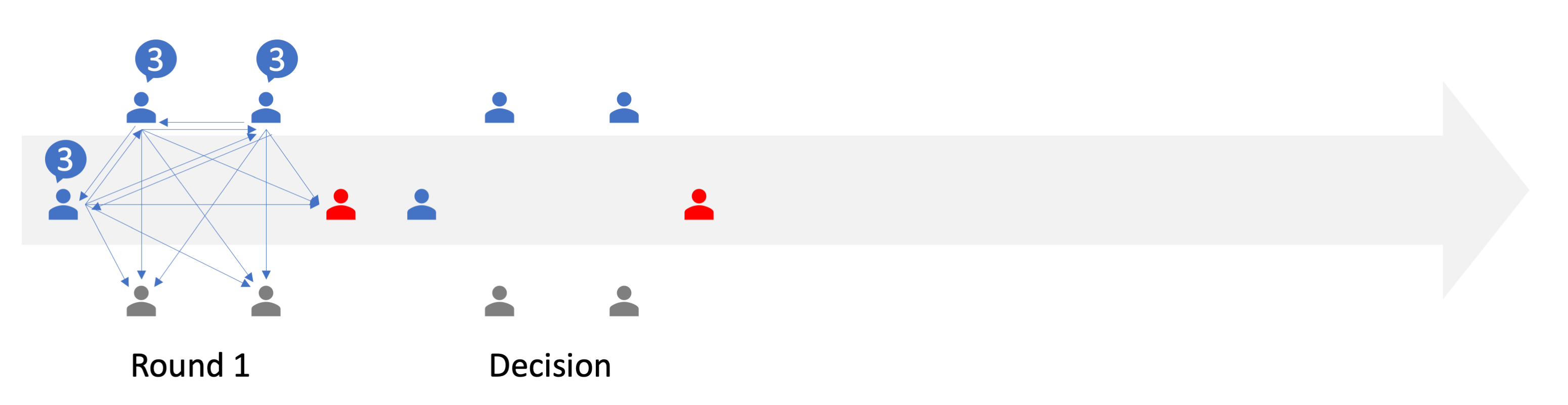} \\
        \hline
        If $r_{tot}=2$ & \includegraphics[width=0.35\linewidth]{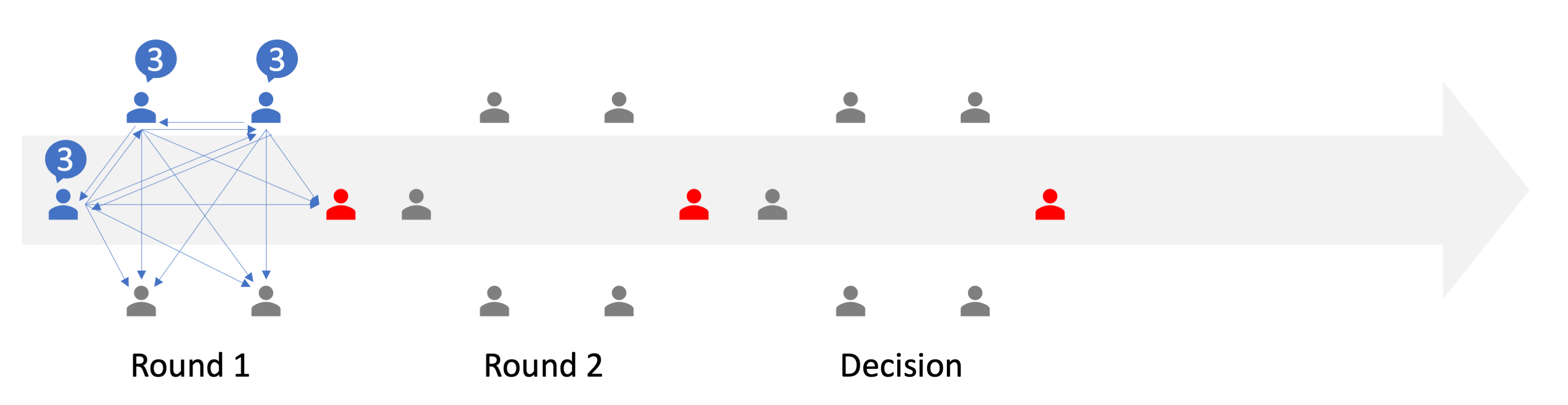} \\
        \hline
        If $r_{tot}=3$ & \includegraphics[width=0.35\linewidth]{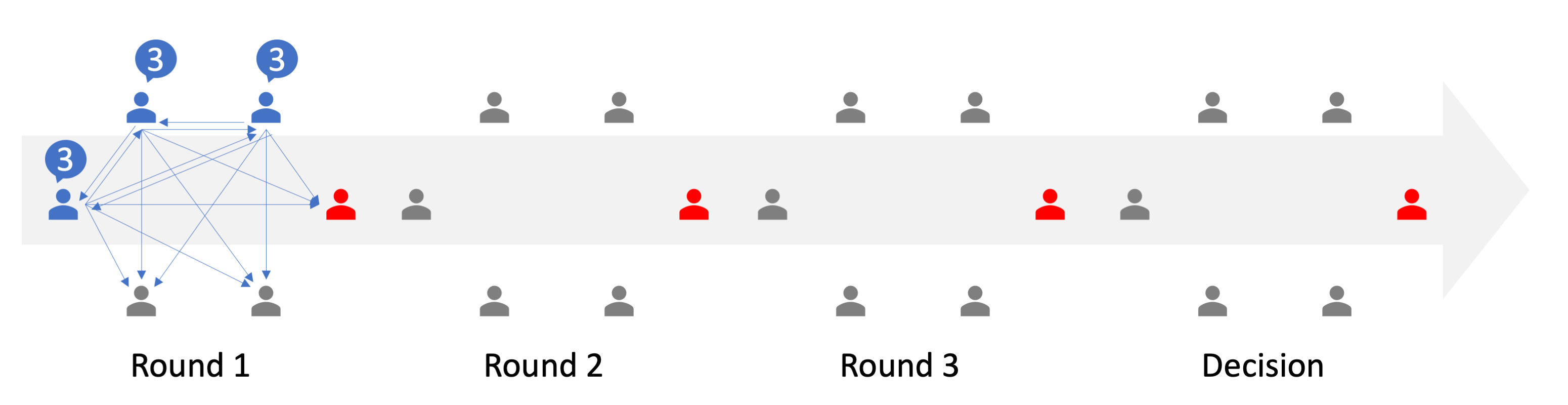} \\
        \hline
    \end{tabular}}\quad
    \subfloat[Attacking strategies that causes the number of active agents $<k+t$ at round 1.]{\begin{tabular}{|c|c|}
        \hline
        If $r_{tot}=1$ & \includegraphics[width=0.35\linewidth]{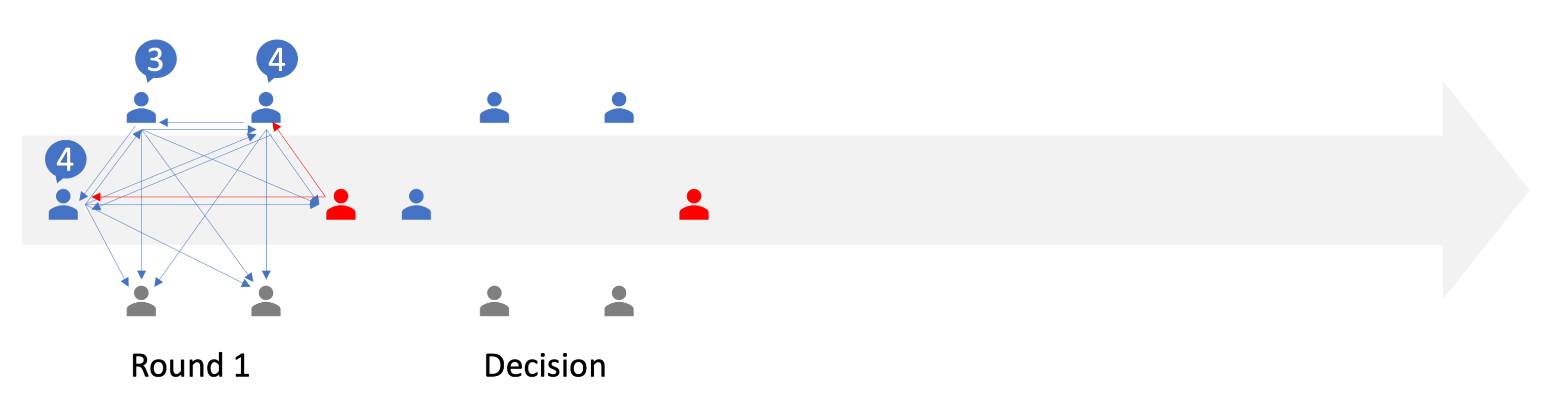} \\
        \hline
        If $r_{tot}=2$ & \includegraphics[width=0.35\linewidth]{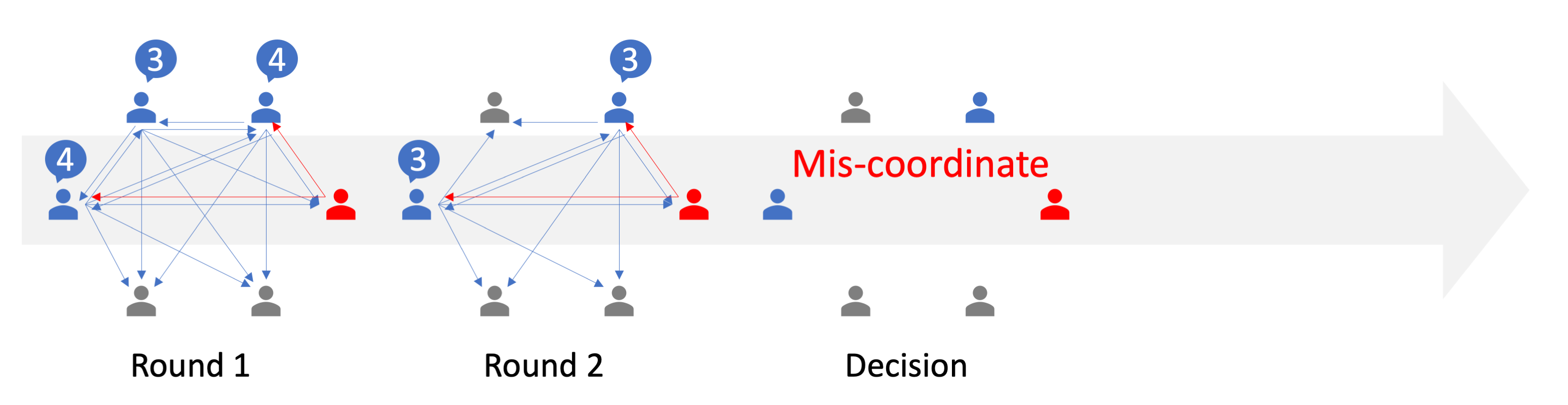} \\
        \hline
        If $r_{tot}=3$ & \includegraphics[width=0.35\linewidth]{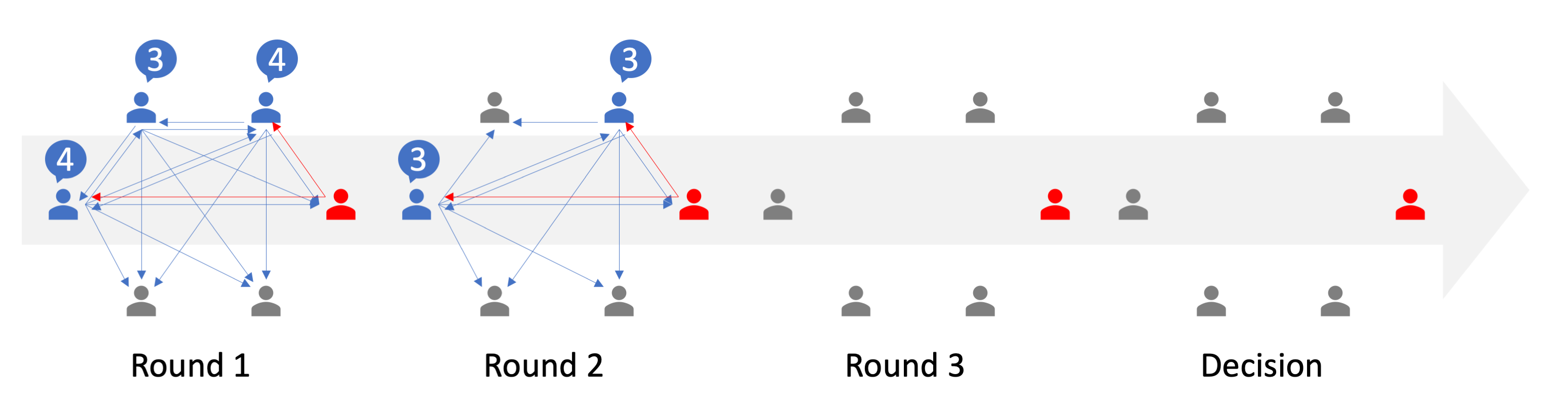} \\
        \hline
    \end{tabular}}\quad
    \subfloat[Attacking strategies that causes the number of active agents $<k+t$ at round 2.]{\begin{tabular}{|c|c|}
        \hline
        If $r_{tot}=1$ & \includegraphics[width=0.35\linewidth]{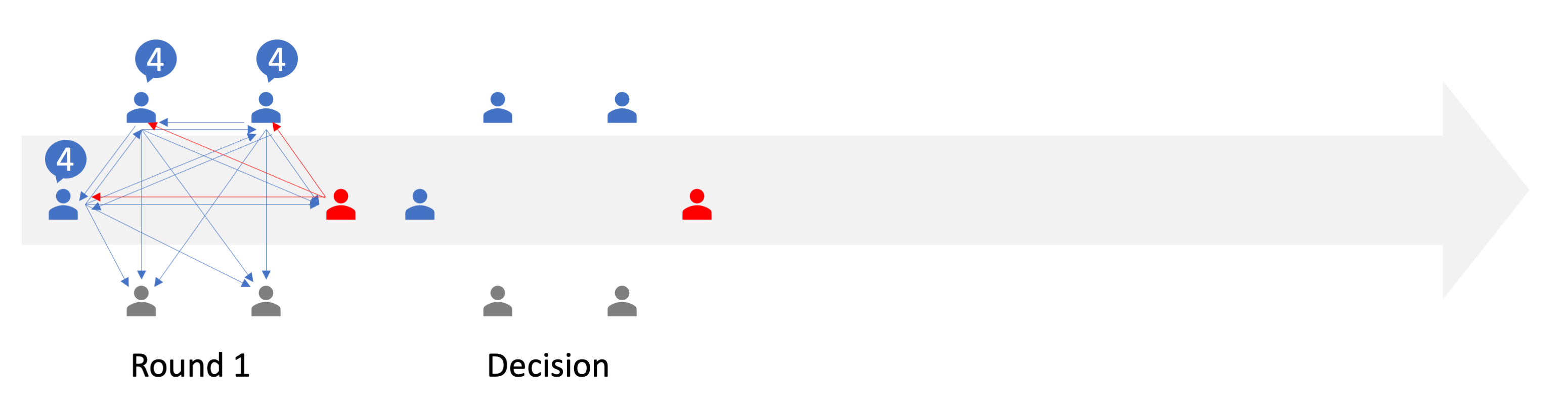} \\
        \hline
        If $r_{tot}=2$ & \includegraphics[width=0.35\linewidth]{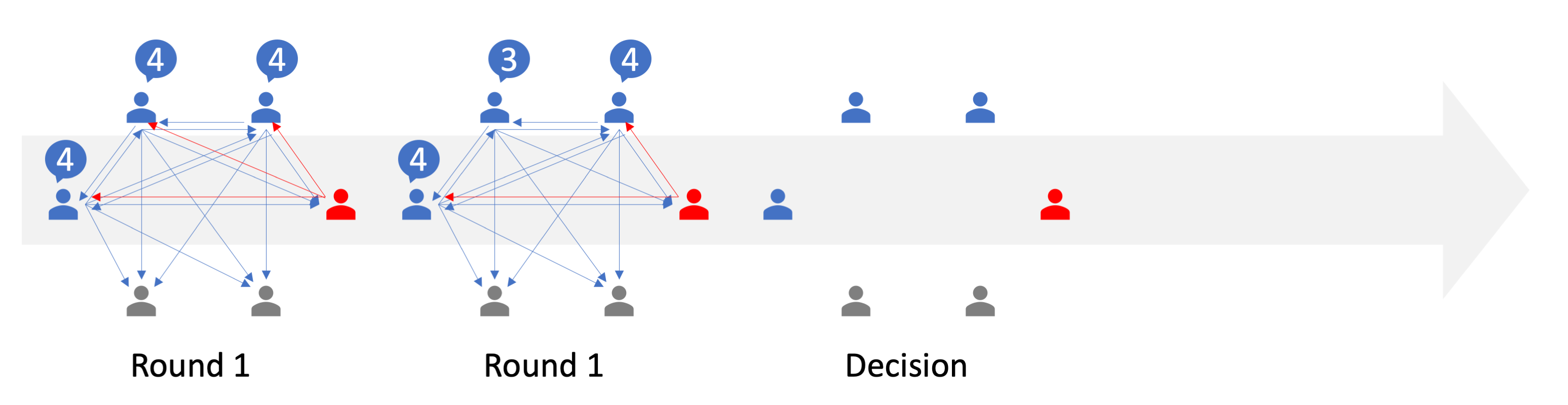} \\
        \hline
        If $r_{tot}=3$ & \includegraphics[width=0.35\linewidth]{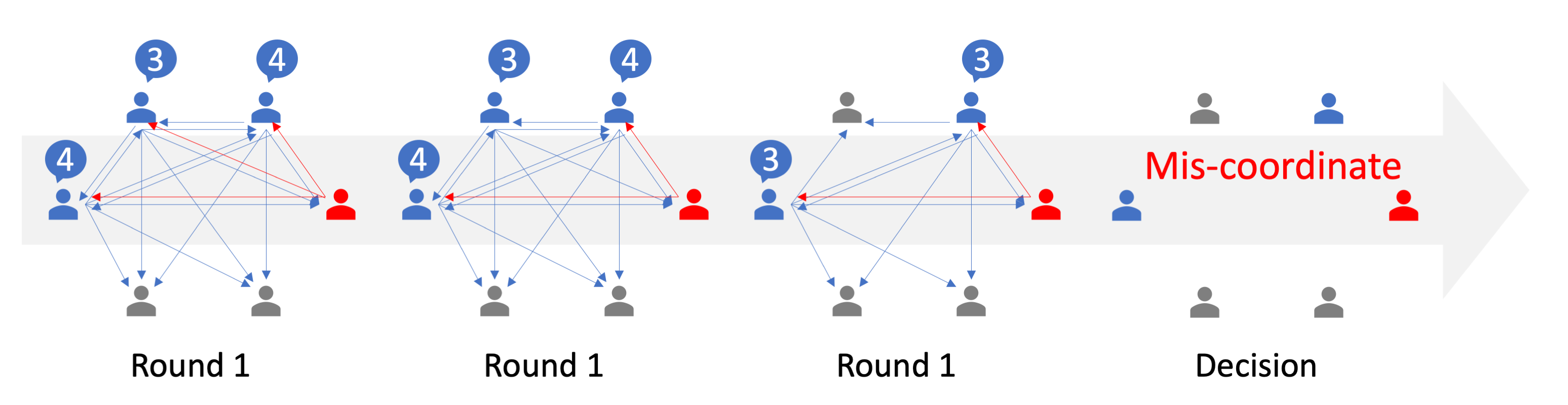} \\
        \hline
    \end{tabular}}\quad
    \subfloat[Attacking strategies that causes the number of active agents $<k+t$ at round 3.]{\begin{tabular}{|c|c|}
        \hline
        If $r_{tot}=1$ & \includegraphics[width=0.35\linewidth]{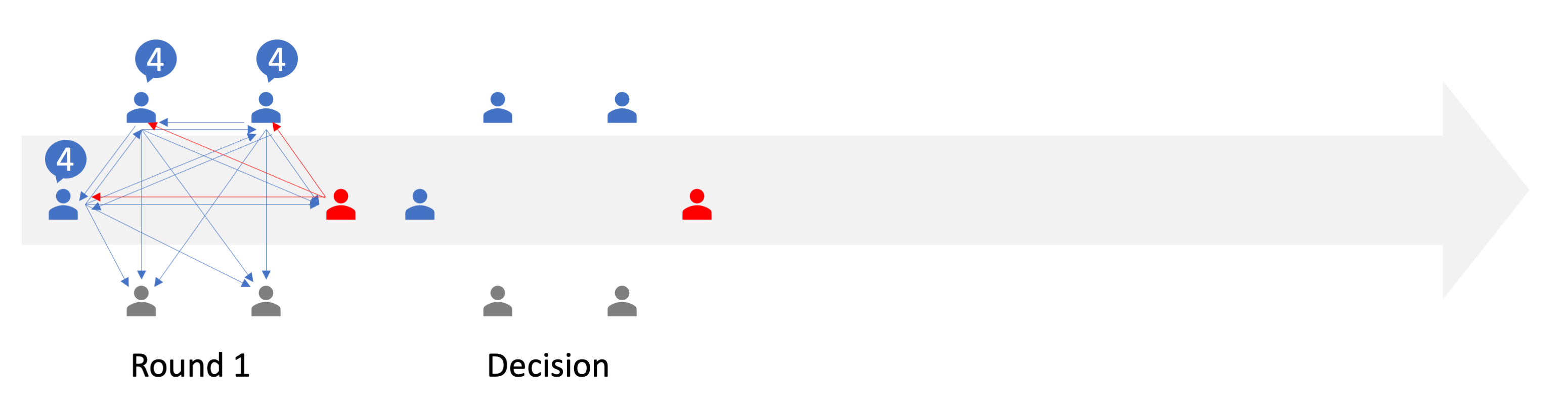} \\
        \hline
        If $r_{tot}=2$ & \includegraphics[width=0.35\linewidth]{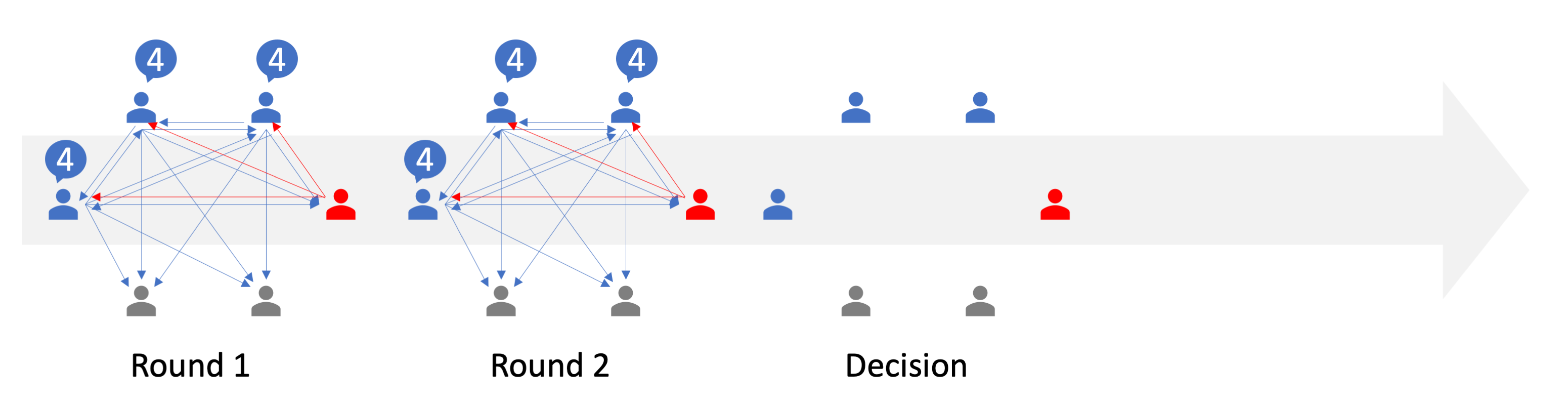} \\
        \hline
        If $r_{tot}=3$ & \includegraphics[width=0.35\linewidth]{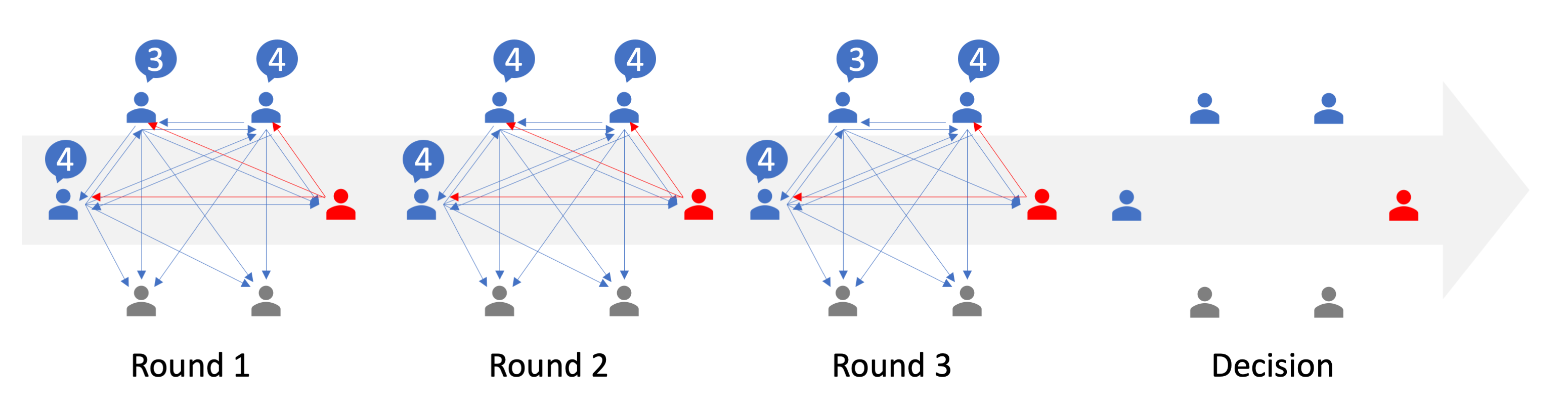} \\
        \hline
    \end{tabular}}
    \caption{The results under all kinds of attacking strategies, with a randomized number of rounds $r_{tot}$.}
    \label{tab:diagram}
\end{table*}

\subsection{Proof of Theorem \ref{thm:protocol}}
\label{ibgp_proof}
\begin{proof}
Before starting the proof, we come up with a lemma.
\begin{lemma}
\label{lemma1}
The number of activating agents $\#(m_{i:}^{(r_0)}=1)$ decreases monotonically when the round $r_0$ increases.
\end{lemma}
Lemma \ref{lemma1} is true because once an agent give up (i.e., $m^{(\cdot)}_{i:}=0$), it will never be activated again by the definition of the protocol.

We prove the robustness of the IBGP Protocol by enumerating all probabilities. In IBGP, there are $3$ types of initialization:
    \begin{enumerate}
        \item $\#(o_i=1)<k$ ($i\in[n]$)
        \item $k\le \#(o_i=1)<k+t$ ($i\in[n]$)
        \item $\#(o_i=1)\ge k+t$ ($i\in[n]$)
    \end{enumerate}

Under the first initialization, in the second round, since $\#(o_i=1)+t<k+t$ (even if all the attackers send $1$ to the agents, it is still lower than the threshold), the messages in the second round $m^{(2)}=0$ definitely. And the final action of each agent must be $a=0$.

Under the third initialization, we prove by induction. If in round $r_0$, the number of activating agents $\#(m^{(r_0)}_{i:}=1)\ge k+t$, in the next round $r_0+1$, the number of activating agents still satisfies $\#(m^{(r_0+1)}_{i:}=1)\ge k+t$. Therefore, since in the first round, the proposition is true, in the decision round the number of activating agents is still greater than $k+t$, and the activating agents will cooperate safely.

Under the second initialization, there's no further conclusion on the number of activating agents in the intermediate rounds. We describe the activation status of each round by Case 1/Case 2 defined as follows.
\begin{itemize}
    \item Case 1: $k\le\#(m^{(r_0)}_{i:}=1)<k+t$
    \item Case 2: $\#(m^{(r_0)}_{i:}=1)\le k$
\end{itemize}
From the Lemma \ref{lemma1}, the rounds switch from Case 1 to Case 2 only once. Denote the round that switches from Case 1 to Case 2 as round $r_{1\rightarrow2}$. We have the following lemma.
\begin{lemma}
    If $r_{1\rightarrow2}$ exists, $\forall r_0>r_{1\rightarrow2},\quad\#(m^{(r_0)}_{i:}=1)=0$.
\end{lemma}
The reason is that since $\#(m^{(r_{1\rightarrow2})}_{i:}=1)\le k$, in the round $r_{1\rightarrow2}+1$, $\#(m^{(r_{1\rightarrow2})}_{i:}=1)=0$.

If the last round $r$ happens in Case 1, all the activating agents would act $a=1$ because no matter what messages the attackers send, $\#(m^{(r_0)}_{i:}=1)+\#\text{attackers}\ge k$. If the last round $r$ happens in Case 2, all the activating agents would act $a=1$ because no matter what messages the attackers send, $\#(m^{(r_0)}_{i:}=1)+\#\text{attackers}\ge k$. If the last round $r>r_{1\rightarrow2}$, since the number of activating agents $\#(m^{(r)}_{i:}=1)=0$, they will act $a=0$ without mis-coordination.

However if $r=r_{1\rightarrow2}$ by coincidence, it is possible that the mis-coordination happens. But since the number of rounds $r$ is randomly drawn by the global randomizer, which is independent of the attackers. The attacker can only guess the final round, and the mis-coordination probability is at most $\max_{r}\{p(r_{tot}=r)\}$.
\end{proof}

\subsection{Multi-agent Environments}

\subsubsection{Robustness in Multi-target Environments}
\label{gen:multi}
\begin{table*}
    \centering
    \begin{tabular}{|c|c|}
        \hline
        Single-target IBGP& Multi-target IBGP\\
        \hline
        $o_i\in\{0,1\},a_i\in\{0,1\}$&$o_i\in\{0,1,2,\cdots,m\},a_i\in\{0,1,2,\cdots,m\}$\\
        \hline
        \makecell{$n$ agents, $1$ target\\coordination threshold $k$}&\makecell{$n$ agents, $m$ targets\\coordination thresholds $k_1,k_2,\cdots,k_m$}\\
        \hline
        $R=\left\{\begin{array}{l}
        \mathbbm{1}(\#(o_i=1,a_i=1)\ge k)\\
        \qquad\qquad\text{if}\ \#(o_i=1)=n\\
        \mathbbm{1}(\#(o_i=1,a_i=1)=0)\\
        \qquad\qquad\text{if}\ \#(o_i=1)<k\\
        \mathbbm{1}(\#(o_i=1,a_i=1)\ge k)\\+\mathbbm{1}(\#(o_i=1,a_i=1)=0)\\
        \qquad\qquad\text{otherwise}\end{array}\right.$&
        $R=\left\{\begin{array}{l}
        \sum_{j\in[m]}\mathbbm{1}(\#(o_i=j,a_i=j)\ge k_j)\\
        \qquad\qquad\text{if}\ \#(o_i=1)=n\\
        \sum_{j\in[m]}\mathbbm{1}(\#(o_i=j,a_i=j)=0)\\
        \qquad\qquad\text{if}\ \#(o_i=1)<k_j\\
        \sum_{j\in[m]}\mathbbm{1}(\#(o_i=j,a_i=j)\ge k_j)\\+\mathbbm{1}(\#(o_i=j,a_i=j)=0)\\
        \qquad\qquad\text{otherwise}\end{array}\right.$\\
        \hline
    \end{tabular}
    \caption{The difference of definition between single-target and multi-target IBGP.}
    \label{tab:multi_ibgp}
\end{table*}
In multi-target environments defined in Table \ref{tab:multi_ibgp}, the IBGP protocol runs independently for each target. If there are $m$ targets, the protocol runs $m$ times for each target, and the final decision is still robust. However, this solution has a drawback: the protocol for each target is as conservative as the single-agent situation (still $(k,t)$-protocol), even if the $m$ is large. This means that $t$ attackers have full influence on each target. Ideally, we want to divide the influence of attackers across multiple targets, as excessive conservativeness would result in low efficiency. To alleviate this problem, we propose adding a new round, called \textit{dispersion defense}, after the first round of the protocol.

\paragraph{Dispersion defense} Acquire a list of $q$ random permutations $p_1,p_2,\cdots,p_q\in[n+t]\rightarrow[n+t]$ from the global randomizer. Each agent defines the received message as $m^{(1)}_{ij}=\mathrm{Majority}_{u\in[q]}\{\tilde{m}^{(1)}_{ip_u(i)}\}$. ($\tilde{m}^{(1)}$ is the original message in the first round.)

The main idea of the dispersion defense is to disarrange the sent messages, making it difficult for attackers to determine the intended receiver. To achieve this, a random permuting process is added after the first round (we call it the proposing round for better understanding). In the dispersion defense, the function $\mathrm{Majority}\{\cdot\}$ returns the majority (more than half) of all elements and returns $0$ if there is no majority.

Apparently, the benign agents broadcast their observations in the first proposing round, and the added dispersion defense doesn't change the messages from benign agents. However, the attackers don't know the permutation $p_1,\cdots,p_q$ when sending messages in the first proposing round and thus cannot adapt the message according to the receiver. As a result, mis-coordination only occurs on a small ratio of targets with high probability, as demonstrated in Theorem \ref{thm1}. Refer to Appendix \ref{proof:thm1} for proof of Theorem \ref{thm1}.
\begin{theorem}
\label{thm1}
    In a $m$-target environment, operating with the $k+\lambda$-protocol will cause at most $\frac{3t}{\lambda}$ targets to mis-coordinate, with probability $1-(m-3)t\cdot k_{max}exp(-2q(\frac{1}{6}-\frac{t}{n+t})^2)$. (Natural assumptions: $f\ll n,t\ll m$)
\end{theorem}

\label{proof:adapt}
\begin{proof}

    First, the main problem of IBGP, the mis-coordination (some valid agents coordinate but other don't), won't happen with high probability due to the structure of IBGP protocol, even if the threshold changes. However, the problem of $\lambda<t$ is that the attackers may mislead agents to coordinate even if there aren't enough valid agents, which we call false coordination in the following proof.

    If $dist(Atk,Atk_{all-1})\ge t-\lambda$, by definition $\forall i\in[n]$,
    \[|\sum_{j=n+1..n+k}m_{Atk,j\rightarrow i}-\sum_{j=n+1..n+k}m_{Atk_{all-1},j\rightarrow i}|\ge t-\lambda,\] which means that \[\sum_{j=n+1..n+k}m_{Atk,j\rightarrow i}\le\lambda.\] If $\#(o_i=1)\le k-1$ (false coordination may happen), the total votes an agent can receive is at most $k-1+\lambda$, so the false coordination never happens.
\end{proof}

\subsubsection{Target Selection in Multi-target environments}
In practical tasks, multiple targets are available for the agents. Luckily, the consensus is still applicable to each target, and thus the key problem is to decide which target to obtain. If the consensus protocol for each target is conducted independently, an agent attending a failed target will not enter the consensus of another target, which is a waste if the second target actually has enough cooperators. We propose a greedy target selection algorithm to handle this problem.

\begin{algorithm}
    \caption{Greedy Target Selection Algorithm}
    \LinesNumbered
    \label{target_selection}
    \KwIn{the number of agents $n$, the number of targets $m$, the rewards of targets $r_1,r_2,\cdots,r_m$, the cooperation threshold $k_1,k_2,\cdots,k_m$, the available set of each target $s_1,s_2,\cdots,s_m\in \{0,1\}^n$}
    \KwOut{selected targets set $T$}
    Sort $m$ targets by the reward $i_1,i_2,\cdots,i_m$ from biggest to smallest\;
    Initialize the occupied agent set $U=\emptyset$\;
    Initialize the selected target set $T=\emptyset$\;
    \For {$j=1\cdots m$}{
        \If{$||s_{i_j}\setminus U||\ge k_{i_j}$}{
            Add $i_j$ to $T$\;
            Add arbitrary $k_{i_j}$ agents from $s_{i_j}\setminus U$ to $T$\;
        }
    }
\end{algorithm}

In fact, the optimal target selection problem is NP-Complete, meaning that there doesn't exist an efficient target selection algorithm even in the centralized setting. Although it is unable to obtain the optimal solution, our Greedy Target Selection Algorithm serves as a $1/k_{max}$-approximation of the optimal solution.
\paragraph{NP-completeness of the target selection problem}
\begin{proof}
We first give a strict definition of the target selection problem.
\paragraph{Target Selection Problem} $n$ agents are pursuing $m$ targets, whose rewards are $r_1,r_2,\cdots,r_m$. The cooperation thresholds are $\theta_1,\theta_2,\cdots,\theta_m$, and the available set of each target is $s_1,s_2,\cdots,s_m\in \{0,1\}^n$. The target selection problem requires $\theta_i$ available agents to cooperate on target $i$ to obtain the reward $r_i$.

We find that the simplified target selection problem is equivalent to the Set Packing problem, which is a classic NP-complete problem.

\paragraph{Simplified Target Selection Problem} $n$ agents are pursuing $m$ targets with the same reward. The cooperation threshold is $\theta$, and the available set of each target is $s_1,s_2,\cdots,s_m\in \{0,1\}^n$.

\paragraph{Set Packing Problem} Suppose there's a finite set $S$ and a list of sets $s_1,s_2,\cdots,s_m$. The set packing problem asks whether there exist $k$ sets that are disjoint.

The simplified target selection problem is just equivalent to the set packing problem, which is NP-complete.
\end{proof}
\paragraph{Proof of $1/k_{max}$-approximation}
\begin{proof}
Denote the selected targets of the greedy algorithm and the optimal algorithm as $S_{greedy}, S_{optimal}$. Consider the greedy selection process from the highest reward to the lowest. When the highest possible target $s_1\in S_{greedy}$ is selected, if $s_1\not\in S_{optimal}$, it can influence at most $k_{max}$ other sets in $S_{optimal}$. But since $s_1$ is of the highest reward, the reward of $s_1$ is at least $1/k$ the reward of all the influenced targets. Similarly, the reward of $s_2$ is also at least $1/k_{max}$ the reward of the influenced targets (the influenced targets aren't double counted). Therefore, the greedy algorithm is a $1/k_{max}$-approximation of the optimal solution. 
\end{proof}

\subsection{Adaption to the Dynamics}
In practical tasks, although the goal is the same, i.e., to keep away from mis-coordination and get rewards, the real rewards and costs could be completely different. As a result, they're basically different problems, however, we could still make use of the nice property of our imperfect consensus protocol, as it is probably safe under any attacks.

The original threshold $k+t$ can be loosened to $k+\lambda$ to be more aggressive, i.e., more likely to succeed with a lower $\#(o_i=1)$, but fail with some probability. But how to choose $\lambda$ is highly related to the environment dynamics, as well as the attackers. To be more detailed, we give some examples below.
\begin{enumerate}
    \item (Environment dynamics) In some environments, the cost is not negative rewards, but the fact that the failing agents would die. It's impossible to value the cost of death directly.
    \item (Environment dynamics) The dynamics is initialized under some probability distribution. The distribution affects the optimal policy. E.g., if $\#(o_i=1)>k$ with high probability, aggressive protocol suffices.
    \item (Attacker) Different attacking policies pose different levels of challenges.
\end{enumerate}
To better illustrate the fact that environment dynamics and attackers would influence the reward and the optimal communicating protocol. We introduce a type of environment and try different thresholds under three kinds of challenging attackers. We also compare the worst success rate under different thresholds. The conclusion is that in practical tasks, the optimal protocol depends on the actual situation.
\begin{table*}[H]
    \centering
    \begin{tabular}{|c|ccc|ccc|ccc|}
        \hline
         ($n,k,f=6,4,2$)&\multicolumn{3}{|c|}{$\lambda=2$}&\multicolumn{3}{|c|}{$\lambda=1$}&\multicolumn{3}{|c|}{$\lambda=0$}  \\
         \hline
         \multirow{2}{*}{success rate}&Atk 1&Atk 2&Atk 3&Atk 1&Atk 2&Atk 3&Atk 1&Atk 2&Atk 3\\
         &0.764&0.997&0.786&0.971&0.746&0.966&0.985&0.644&0.998\\
         \hline
         worst success rate&\multicolumn{3}{|c|}{\textbf{0.764}}&\multicolumn{3}{|c|}{0.746}&\multicolumn{3}{|c|}{0.644}\\
         \hline
    \end{tabular}
    \begin{tabular}{|c|ccc|ccc|ccc|}
        \hline
         ($n,k,f=10,4,2$)&\multicolumn{3}{|c|}{$\lambda=2$}&\multicolumn{3}{|c|}{$\lambda=1$}&\multicolumn{3}{|c|}{$\lambda=0$}  \\
         \hline
         \multirow{2}{*}{success rate}&Atk 1&Atk 2&Atk 3&Atk 1&Atk 2&Atk 3&Atk 1&Atk 2&Atk 3\\
         &0.792&0.995&0.8&0.964&0.847&0.969&0.992&0.763&0.994\\
         \hline
         worst success rate&\multicolumn{3}{|c|}{0.792}&\multicolumn{3}{|c|}{\textbf{0.847}}&\multicolumn{3}{|c|}{0.763}\\
         \hline
    \end{tabular}
    \caption{This environment is just IBGP extended to 10 steps. There is still one target, which requires $k$ agents to cooperate simultaneously. But the target can only be obtained once. After the target is obtained, the game ends. Any mis-coordinating agents would die. The observations is initialized as $P(\#(o_i=1)=0)=P(\#(o_i=1)=1)=\cdots=P(\#(o_i=1)=6)=\frac{1}{7}$. The upper and lower table shows the results with $n=6$ and $n=10$. The three attackers are all-1-all-0 attacker, all-1 attacker and all-0 attacker. A detailed description of attackers is in the following part.}
    \label{tab:exp}
\end{table*}


\subsubsection{Adaption to the Attacker}
\begin{table}[H]
    \centering
    \begin{tabular}{|c|c|c|c|}
        \hline
         &$\lambda=0$&$\lambda=1$&$\lambda=2$  \\
         \hline
         $p=0$&\textbf{0.460}&0.312&0.162\\
         $p=0.1$&\textbf{0.462}&0.308&0.159\\
         $p=0.2$&\textbf{0.461}&0.310&0.163\\
         $p=0.3$&\textbf{0.459}&0.314&0.162\\
         $p=0.4$&\textbf{0.459}&0.316&0.166\\
         $p=0.5$&\textbf{0.456}&0.331&0.183\\
         $p=0.6$&\textbf{0.448}&0.361&0.220\\
         $p=0.7$&\textbf{0.445}&0.404&0.266\\
         $p=0.8$&0.441&\textbf{0.444}&0.304\\
         $p=0.9$&0.434&\textbf{0.454}&0.352\\
         $p=1$&0.433&0.437&\textbf{0.461}\\
         \hline
    \end{tabular}
    \caption{Taking different $\lambda$s when the probability of the random attacker differs.}
    \label{tab:adp_atk}
\end{table}
The optimal $\lambda$ also varies when the attacker changes. For example, according to Table \ref{tab:adp_atk}, when the attacker is a random attacker with probability $0,0.1,0.2,\cdots,1$, the optimal $\lambda$ varies.

In practical scenarios, the attacker may follow a distribution that does not pose a significant threat to our consensus. Therefore, we can adjust the network's hidden layer parameters. More specifically, we relax $\lambda$ to some value smaller than $t$. This change reduces the required number of available agents $\#(o_i=1)$, increasing the likelihood of success. We provide a guarantee (Theorem \ref{thm:adapt}) for the relaxed protocol, assuming certain conditions about the attackers. To begin, we define the distance between two attacking schemes, $dist(Atk_1, Atk_2)=\min_{env}\min_{i\in[n]}\{|\sum_{j=n+1..n+k}m_{Atk_1,j\rightarrow i}-\\\sum_{j=n+1..n+k}m_{Atk_2,j\rightarrow i}|\}$.

\begin{theorem}
\label{thm:adapt}
    If $dist(Atk, Atk_{all-1})\ge t-\lambda$, the new protocol with a threshold $k+\lambda$ would not fail with high probability. ($Atk_{all-1}$ is an attacker that sends $1$ all the time.) 
\end{theorem}

\label{proof:thm1}
\begin{proof}
For clarity, we focus only on the number of influenced targets in the proof. The small probability of the consensus failure keeps the same, so we don't cover this part in our analysis.

    In the first proposing round, each attacker $i$ send messages $\tilde{m}^{(1)}_{i:}$ where there are $|U|$ targets s.t. $\forall u\in U,\#(\tilde{m}_{i?}=u)\ge\frac{n+t}{3}$ and $m-|U|$ targets s.t. $\forall u\not\in U,\#(\tilde{m}_{i?}=u)<\frac{n+t}{3}$. Since there are $n+t$ agents in total, $|U|\le (n+t)/\frac{n+t}{3}=3$.

    
    For each agent $j$, denote $\#(\tilde{m}_{i?}=o_j)=\tau$. Under such $\tau$,
    \begin{align*}
        P(m^{(1)}_{ij}=o_j)&=P(\mathrm{Majority}_{u\in[q]}\{\tilde{m}^{(1)}_{ip_u(i)}\}=o_j)\\
        &\le P(\#(\tilde{m}^{(1)}_{ip_u(i)}=o_j)\ge \frac{q}{2})\\
        &=P_{Binomial(q,\frac{\tau+t}{n+t})}(X\ge\frac{q}{2})\\
    \end{align*}
    If $\frac{\tau+t}{n+t}<\frac{1}{2}$, $P(m^{(1)}_{ij}=o_j)\le exp(-2q(\frac{1}{2}-\frac{\tau+t}{n+t})^2)$ because of Hoeffding Inequality.
    
    $P(m^{(1)}_{ij}=o_j)$ is understood as the probability that agent $j$ is influenced by attacker $j$ under $\tau$. For a target $u$, $P(\exists j\in[n], m^{(1)}_{ij}=o_j=u)\le k_{max}P(m^{(1)}_{ij}=o_j)$ is understood as the probability that target $u$ is influenced by attacker $j$.

    If $o_j\not\in U$, $P(\exists j\in[n], m^{(1)}_{ij}=o_j=u)\le k_{max}exp(-2q(\frac{1}{2}-\frac{\frac{n+t}{3}+t}{n+t})^2)\equiv\epsilon$.

    If a target mis-coordinate, there must be more than $\lambda$ attackers influencing it. Therefore, if more than $\frac{3t}{\lambda}$ targets mis-coordinate, there must be at least $\frac{3t}{\lambda}\cdot\lambda$ influencing pairs. (The influence pair $(i,j)\in[t]\times[m]$ means that attacker $i$ influence target $j$.) But the probability upper bound of all influencing pairs is $\{\underbrace{1,1,\cdots,1}_{3t},\underbrace{\epsilon,\epsilon,\cdots,\epsilon}_{(m-3)t}\}$. So the probablity of more than $\frac{3t}{\lambda}$ mis-coordination is less than $(m-3)t\epsilon$.
\end{proof}
    
    This conclusion also generalizes to varying $\lambda$ values for each agent in Appendix \ref{apd:lambda}.

\subsubsection{Varying $\lambda$ for each agent}
\label{apd:lambda}
When different agents have a different $\lambda_i$, the protocol is as follows.
\begin{enumerate}
    \item The global randomizer initializes the number of rounds $r$ from a distribution $r\sim\mathcal{R}$
    \item First round: Each agent broadcasts its observation $m^{(1)}_{i:}=o_i$
    \item Round $r_{now}\in\{2\cdots r\}$: Each agent $i\in\{i|m^{(r_{now})}_i=1\}$ broadcasts $m^{(2)}_{i:}=\begin{cases}
    0&k\le\sum_{j\in[n+t]}\mathbbm{1}(m^{(1)}_{ji}=1)< k+\lambda_i\\
    0&\sum_{j\in[n+t]}\mathbbm{1}(m^{(1)}_{ji}=1)< k\\
    1&\sum_{j\in[n+t]}\mathbbm{1}(m^{(1)}_{ji}=1)\ge k+\lambda_i
    \end{cases}$
    \item Decision making round: Each agent $i\in\{i|m^{(r_{now})}_i=1\}$ select action $a_i=\begin{cases}
    1&k\le\sum_{j\in[n+t]}\mathbbm{1}(m^{(1)}_{ji}=1)< k+\lambda_i\\
    0&\sum_{j\in[n+t]}\mathbbm{1}(m^{(1)}_{ji}=1)< k\\
    1&\sum_{j\in[n+t]}\mathbbm{1}(m^{(1)}_{ji}=1)\ge k+\lambda_i
    \end{cases}$
\end{enumerate}

\begin{table}[H]
    \centering
    \begin{tabular}{|c|c|c|c|c|}
        \toprule
         ($n,k,t=6,4,2$)&$\lambda=0$&$\lambda=1$&$\lambda=2$&varying $\lambda$ \\
         \midrule
         $p=0$&\textbf{0.052}&0.005&0.000&0.000\\
         $p=0.1$&\textbf{0.051}&0.005&0.000&0.000\\
         $p=0.2$&\textbf{0.051}&0.005&0.000&0.001\\
         $p=0.3$&\textbf{0.051}&0.005&0.000&0.002\\
         $p=0.4$&\textbf{0.047}&0.005&0.000&0.000\\
         $p=0.5$&\textbf{0.036}&0.005&0.000&0.000\\
         $p=0.6$&\textbf{0.013}&0.008&0.000&0.002\\
         $p=0.7$&-0.026&\textbf{0.020}&0.002&0.005\\
         $p=0.8$&-0.068&\textbf{0.030}&0.004&0.008\\
         $p=0.9$&-0.109&\textbf{0.031}&0.013&0.023\\
         $p=1$&-0.191&-0.087&\textbf{0.052}&0.050\\
         \midrule
         Avg&-0.008&0.002&0.006&\textbf{0.008}\\
         \bottomrule
    \end{tabular}
    \caption{Optimal $\lambda$ setting for different attacker settings.}
    \label{tab:adp_atk_new}
\end{table}

For each agent $i_0$, we define the distance between two attacking schemes, $dist(Atk_1,Atk_2)=\min_{env}\{\sum_{j=n+1..n+k}m_{Atk_1,j\rightarrow i_0}-\sum_{j=n+1..n+k}m_{Atk_2,j\rightarrow i_0}\}$. If $\forall i\in[n],dist(Atk_{i},Atk_{all-1})\ge t-\lambda_i$, the new protocol with thresholds $k+\lambda_1,k+\lambda_2,\cdots,k+\lambda_n$ would not fail with high probability. ($Atk_{all-1}$ is an attacker that sends $1$ all the time.)

In Table \ref{tab:adp_atk_new}, we compare the performance of using the same $\lambda$ and varying $\lambda$ to deal with random attackers with various probabilities. We use a single-target task and measure the average reward in a single step. We see that varying $\lambda$ leads to the most adaptive protocol on average, indicating the benefit of microscopic complexity.


\subsection{Implementation details}
\label{apd:impl}
\subsubsection{Algorithms}
Since this paper focus on the communicative attacks and also communicative defense, we build our algorithms based on the codebase in \cite{tonghan2020learning}. For simplicity, we remove all the losses on communication learning except the downstream loss, making the codebase only a multi-agent QMIX algorithm with learnable communications. 

The first baseline AME \cite{sun2023certifiably} publishes its code, but is the continuous-action version. Therefore, we re-implement AME in our code for fair comparison. The second baseline \cite{xue2022mis} is not open-source, and one of its contributions is based on an impractical assumption that the disturbance on messages is only a small noise. So we implement recursive training, which is its second contribution, to compare in the experiment section.

Our algorithm and AME includes two phases, the training and testing phase. In the experiment, the testing phase means that we freeze the agents' policy and communication and train the attackers to simulate the real attacking scheme. The training phase lasts for $t_{train}$ steps and the testing phase (the attacker learning phase) lasts for $t_{test}$ steps. And in the recursive training, we recursively train the agents and attackers with a fixed interval of steps $t$.
\begin{table*}[htbp]
    \centering
    \begin{tabular}{|c|ccc|c|ccc|}
        \hline
        &\multicolumn{3}{|c|}{IBGP Protocol}&Recursive training&\multicolumn{3}{|c|}{AME}\\
        \hline
        &$t_{train}$&$t_{test}$&$r_{max}$&$t$&$t_{train}$&$t_{test}$&$k_{\text{AME}}$\\
        \hline
         Predator-prey$(4,1,2,1)$&$1e7$&$1e7$&$5$&$1e7$&$5e6$&$5e6$&$1$\\
        \hline
         Predator-prey$(5,2,2,1)$&$1e7$&$1e7$&$5$&$5e6$&$5e6$&$5e6$&$1$\\
        \hline
         Predator-prey$(20,4,2,2)$&$8e6$&$4e6$&$5$&$1e7$&$5e6$&$5e6$&$4$\\
        \hline
         Predator-prey$(20,1,4,2)$&$4e6$&$4e6$&$5$&$3e6$&$5e6$&$5e6$&$4$\\
        \hline
         Hallway$(3,1,2,1)$&$1e7$&$1e7$&$5$&$3e6$&$5e6$&$5e6$&$1$\\
        \hline
         Hallway$(10,1,5,2)$&$5e5$&$5e5$&$5$&$5e6$&$5e6$&$5e6$&$4$\\
        \hline
         4bane\_vs\_1hM$(4,1,3,1)$&$1e7$&$1e7$&$5$&$3e6$&$5e6$&$5e6$&$1$\\
        \hline
         3z\_vs\_1r$(3,1,2,1)$&$4e6$&$4e6$&$5$&$5e6$&$5e6$&$5e6$&$1$\\
        \hline
    \end{tabular}
    \caption{The training/testing steps and hyperparameters used in the experiments.}
    \label{apd:hyper}
\end{table*}

The training/testing steps and hyperparameters are shown in Table \ref{apd:hyper}. The steps aren't tunable parameters, as we adopt the proper step where the learning process is converged. $r_{max}$ is the maximum rounds in IBGP Protocol, and the $k_{\text{AME}}$ is the built-in hyperparameter in AME algorithms.

There's no exact requirement of computing sources, because the algorithm and environment are not computationally intensive.

\subsubsection{Environment}
We make modifications to the predator-prey environment to make it easier to learn in the training phase. Specifically, we use fixed preys so agents can coordinate in an easier pattern. We use the map size $5$ in the main results and map size $10$ in the adaption process, because the map size $10$ is more challenging and the $k+t$ protocol will not $100\%$ succeed and has a potential to improve in the adaption process. We also add an auxiliary reward based on the distance to the target to encourage exploration in Predator-prey$(4,1,2,1)$ and Predator-prey$(5,2,2,1)$ to make the training phase converge to a good solution.

\begin{figure}[H]
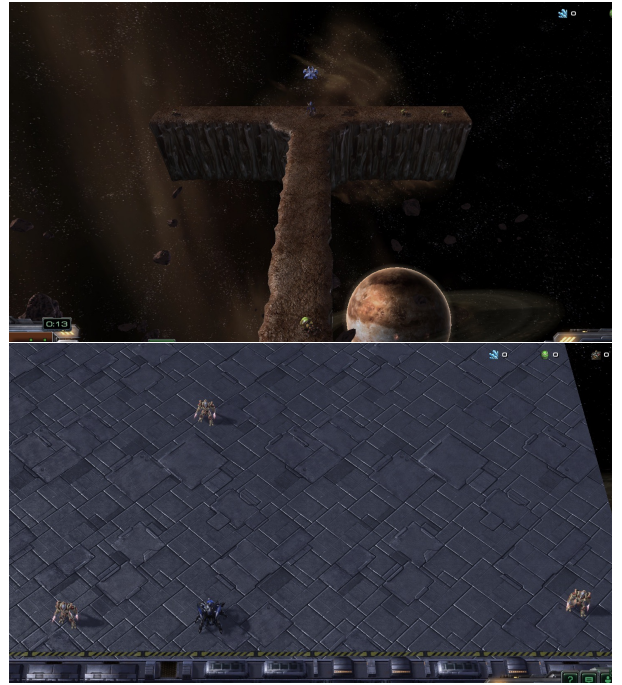

\centering
    \includegraphics[width=0.45\textwidth]{images/bane.jpeg}
    \includegraphics[width=0.45\textwidth]{images/3z.jpeg}
    \caption{The screenshot of 2 StarCraft II enironments.}
    \label{fig:sc2}
\end{figure}
We attach the figures of the StarCraft II environment 4bane\_vs\_1hM and 3z\_vs\_1r in Figure \ref{fig:sc2}.

\subsubsection{Training-testing phase details} 
\begin{figure*}[!h]
\centering
\includegraphics[width=\textwidth]{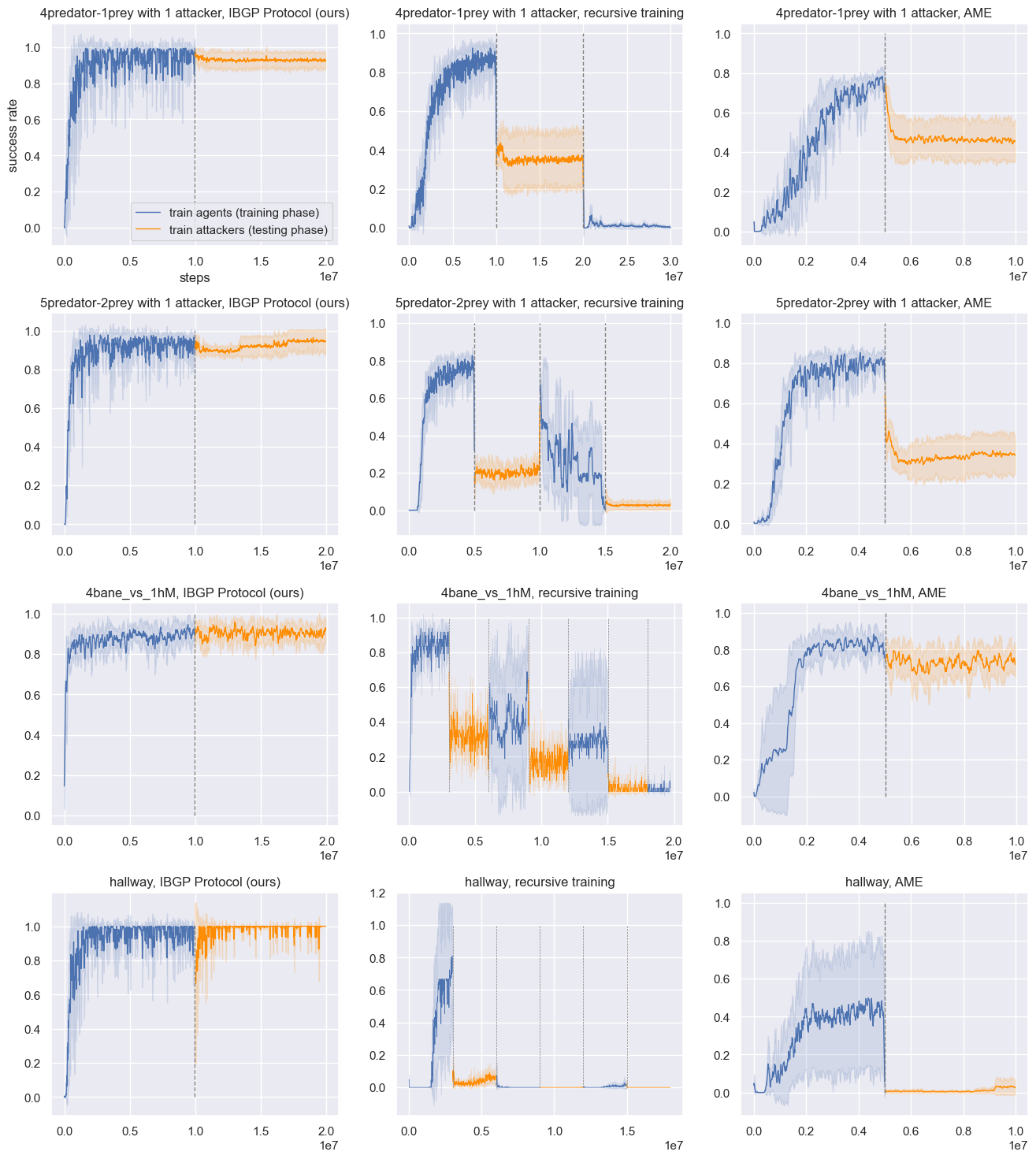}
\caption{The learning curves of four environments, comparing IBGP Protocol, recursive training and AME.}
\label{main_curve}
\end{figure*}

To better illustrate how we evaluate the robustness of algorithms, we include the learning curves (Figure \ref{main_curve}) of four environments in the main results (Section \ref{exp:main}). To compare the zero-shot robustness of algorithms, we display the experiment results by coloring the curve in the training phase in blue and coloring the curve in the testing phase in orange. Also, the testing phase is actually the training process of the attackers. Therefore, if the algorithm isn't robust enough against communicative attacks, the learning curve in the testing phase would decay significantly.

The left column shows the learning curve of the IBGP protocol. And the middle column is the learning curve of recursive training, meaning that the training process of agents and attackers is repeated. Recursive training is one of the contributions of the algorithm in \cite{xue2022mis}. We also compare the AME algorithm in \cite{sun2023certifiably} shown in the right column.

From the learning curves in Figure \ref{main_curve}, the IBGP Protocol maintains the performance from the training phase to the testing phase. But the recursive training fails to be robust when switched to the testing phase. The AME algorithm performs well in part of the environments in the testing phase but still endures a performance decay in the testing phase.

\subsubsection{Details of the Sensor Network problem}

\begin{figure*}
    \centering
    \includegraphics[width=0.7\linewidth]{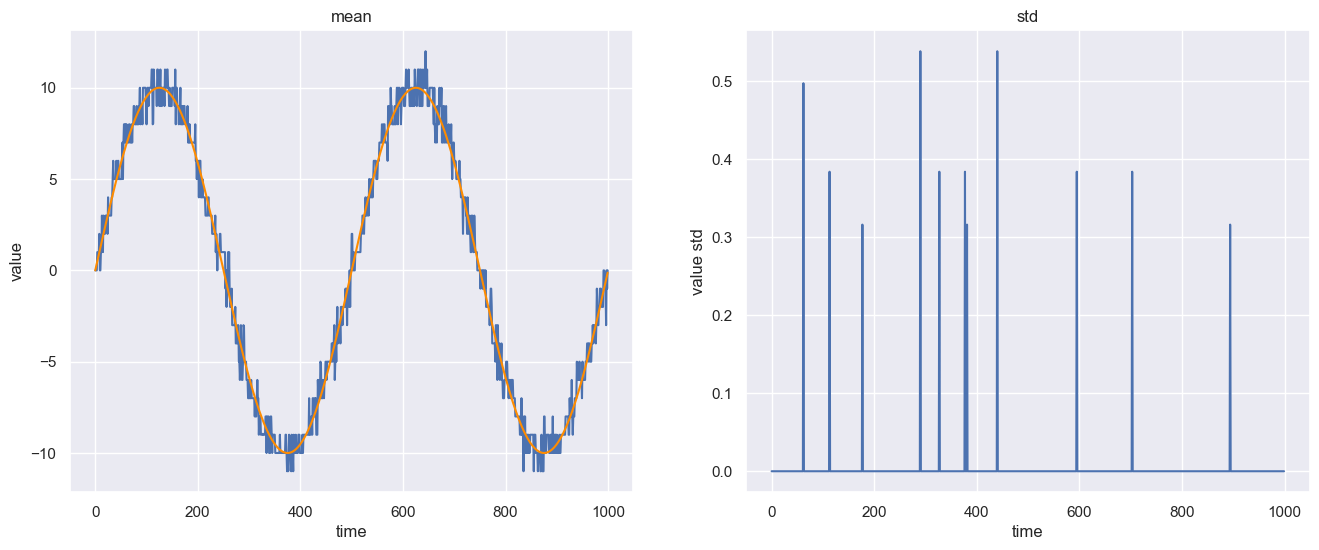}
    \caption{The simulation result including mean and standard deviation of Sensor Networks Problem based on IBGP Protocol.}
    \label{fig:sensor_mean_std}
\end{figure*}
In Figure \ref{fig:sensor_mean_std}, the average belief of all agents is consistent with the true signal, and the standard deviation is nearly zero, indicating that all agents share the same belief.

The correctness of the algorithm is because the following theorem.
\begin{theorem}
\label{thm:sensor}
    If there is at most one attacker in any neighborhood that holds consensus, the consensus-based algorithm is robust to broadcast the signal. (Assumptions: at least two adjacent agents observe the new signal; each neighborhood contains more than three agents.)
\end{theorem}
\label{apd:sensor}
The proof of Theorem \ref{thm:sensor} is as follows.
\begin{proof}
Only the correct signal can pass the consensus, because of the robustness of the protocol. And each agent is updated only once, so the algorithm will finally terminate at each timestep $t$. Due to the connectivity of the sensor network, the new signal will also be broadcasted to all the agents.
\end{proof}

\begin{figure*}[!h]
    \centering
    \includegraphics[width=0.9\textwidth]{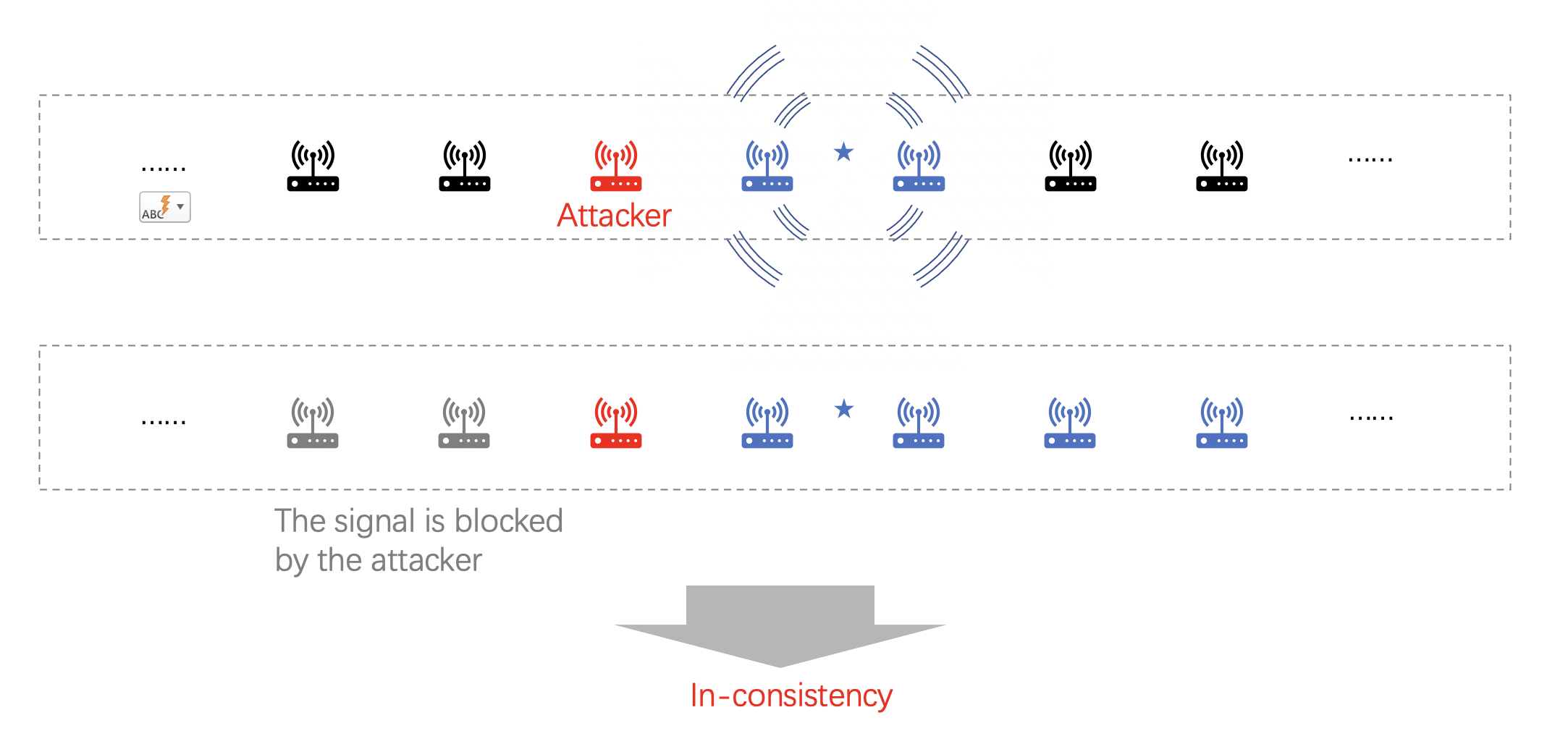}
    \caption{One counterexample of the vanilla broadcasting.}
    \label{fig:sensor_counter}
\end{figure*}
\paragraph{An counterexample without the consensus-based algorithm} To better illustrate why the consensus protocol is essential in such Sensor Network problems, we give an counterexample where vanilla broadcasting is used to transfer the signal in Figure \ref{fig:sensor_counter}. In this case, the sensor network is in a 1-dimension array. Once a new target is observed at some position, the sensors with new observations (blue) try to broadcast the new signal of target position. However, the existence of the attacker sends fake messages and may block the broadcasting, causing in-consistency of the whole system.

\subsection{Detailed description about Figure \ref{fig:intro}}
\label{intro_fig_detail}
Denote the whole number of agents as $n$, the percentage of malicious agents as $r$, and the coordination threshold as $k$. The x-axis and y-axis are $r$ and $n(1-r)-k$. The requirement of BGP is $n>3(1-r)+1$ due to the $n>3t+1$ requirement. The requirement of IBGP is $n*r>k+n*(1-r)$ due to the requirement that the number of good agents $>k+t$.

\end{document}